\documentclass[12pt]{article}

\usepackage{etex}

\usepackage{fullpage}

\usepackage{tikz}
\usetikzlibrary{decorations,shapes} 

\usepackage{tikz-cd}
\tikzcdset{
  shift left/.default=+0.7ex,% default: +0.56ex
  shift right/.default=+0.7ex}

\newenvironment{tikzar}[1][]{{}\kern-4pt\begin{tikzcd}[ampersand replacement=\&,#1]}%
{\end{tikzcd}\kern-4pt{}}

\usepackage{enumitem}
\setlist[itemize]{noitemsep, topsep=0pt}

\usepackage{amsmath,stmaryrd}
\usepackage{amsfonts,amssymb}%,txfonts,
\usepackage{pstricks}
\usepackage{amsthm}
\usepackage{rotating}

\usepackage[LGR,T1]{fontenc}
\newcommand{\textgreek}[1]{\begingroup\fontencoding{LGR}\selectfont#1\endgroup}

\newdimen\proofrulebreadth \proofrulebreadth=.05em
\newdimen\proofdotseparation \proofdotseparation=1.25ex
\newdimen\proofrulebaseline \proofrulebaseline=2ex
\newcount\proofdotnumber \proofdotnumber=3
\let\then\relax
\def\hfi{\hskip0pt plus.0001fil}
\mathchardef\squigto="3A3B
\newif\ifinsideprooftree\insideprooftreefalse
\newif\ifonleftofproofrule\onleftofproofrulefalse
\newif\ifproofdots\proofdotsfalse
\newif\ifdoubleproof\doubleprooffalse
\let\wereinproofbit\relax
\newdimen\shortenproofleft
\newdimen\shortenproofright
\newdimen\proofbelowshift
\newbox\proofabove
\newbox\proofbelow
\newbox\proofrulename
\def\shiftproofbelow{\let\next\relax\afterassignment\setshiftproofbelow\dimen0 }
\def\shiftproofbelowneg{\def\next{\multiply\dimen0 by-1 }%
\afterassignment\setshiftproofbelow\dimen0 }
\def\setshiftproofbelow{\next\proofbelowshift=\dimen0 }
\def\setproofrulebreadth{\proofrulebreadth}

\def\prooftree{% NESTED ZERO (\ifonleftofproofrule)
\ifnum  \lastpenalty=1
\then   \unpenalty
\else   \onleftofproofrulefalse
\fi
\ifonleftofproofrule
\else   \ifinsideprooftree
        \then   \hskip.5em plus1fil
        \fi
\fi
\bgroup% NESTED ONE (\proofbelow, \proofrulename, \proofabove,
\setbox\proofbelow=\hbox{}\setbox\proofrulename=\hbox{}%
\let\justifies\proofover\let\leadsto\proofoverdots\let\Justifies\proofoverdbl
\let\using\proofusing\let\[\prooftree
\ifinsideprooftree\let\]\endprooftree\fi
\proofdotsfalse\doubleprooffalse
\let\thickness\setproofrulebreadth
\let\shiftright\shiftproofbelow \let\shift\shiftproofbelow
\let\shiftleft\shiftproofbelowneg
\let\ifwasinsideprooftree\ifinsideprooftree
\insideprooftreetrue
\setbox\proofabove=\hbox\bgroup$\displaystyle % NESTED TWO
\let\wereinproofbit\prooftree
\shortenproofleft=0pt \shortenproofright=0pt \proofbelowshift=0pt
\onleftofproofruletrue\penalty1
}

\def\eproofbit{% NESTED TWO
\ifx    \wereinproofbit\prooftree
\then   \ifcase \lastpenalty
        \then   \shortenproofright=0pt  % 0: some other object, no indentation
        \or     \unpenalty\hfil         % 1: empty hypotheses, just glue
        \or     \unpenalty\unskip       % 2: just had a tree, remove glue
        \else   \shortenproofright=0pt  % eh?
        \fi
\fi
\global\dimen0=\shortenproofleft
\global\dimen1=\shortenproofright
\global\dimen2=\proofrulebreadth
\global\dimen3=\proofbelowshift
\global\dimen4=\proofdotseparation
\global\count255=\proofdotnumber
$\egroup  % NESTED ONE
\shortenproofleft=\dimen0
\shortenproofright=\dimen1
\proofrulebreadth=\dimen2
\proofbelowshift=\dimen3
\proofdotseparation=\dimen4
\proofdotnumber=\count255
}

\def\proofover{% NESTED TWO
\eproofbit % NESTED ONE
\setbox\proofbelow=\hbox\bgroup % NESTED TWO
\let\wereinproofbit\proofover
$\displaystyle
}%
\def\proofoverdbl{% NESTED TWO
\eproofbit % NESTED ONE
\doubleprooftrue
\setbox\proofbelow=\hbox\bgroup % NESTED TWO
\let\wereinproofbit\proofoverdbl
$\displaystyle
}%
\def\proofoverdots{% NESTED TWO
\eproofbit % NESTED ONE
\proofdotstrue
\setbox\proofbelow=\hbox\bgroup % NESTED TWO
\let\wereinproofbit\proofoverdots
$\displaystyle
}%
\def\proofusing{% NESTED TWO
\eproofbit % NESTED ONE
\setbox\proofrulename=\hbox\bgroup % NESTED TWO
\let\wereinproofbit\proofusing
\kern0.3em$
}

\def\endprooftree{% NESTED TWO
\eproofbit % NESTED ONE
  \dimen5 =0pt% spread of hypotheses
\dimen0=\wd\proofabove \advance\dimen0-\shortenproofleft
\advance\dimen0-\shortenproofright
\dimen1=.5\dimen0 \advance\dimen1-.5\wd\proofbelow
\dimen4=\dimen1
\advance\dimen1\proofbelowshift \advance\dimen4-\proofbelowshift
\ifdim  \dimen1<0pt
\then   \advance\shortenproofleft\dimen1
        \advance\dimen0-\dimen1
        \dimen1=0pt
        \ifdim  \shortenproofleft<0pt
        \then   \setbox\proofabove=\hbox{%
                        \kern-\shortenproofleft\unhbox\proofabove}%
                \shortenproofleft=0pt
        \fi
\fi
\ifdim  \dimen4<0pt
\then   \advance\shortenproofright\dimen4
        \advance\dimen0-\dimen4
        \dimen4=0pt
\fi
\ifdim  \shortenproofright<\wd\proofrulename
\then   \shortenproofright=\wd\proofrulename
\fi
\dimen2=\shortenproofleft \advance\dimen2 by\dimen1
\dimen3=\shortenproofright\advance\dimen3 by\dimen4
\ifproofdots
\then
        \dimen6=\shortenproofleft \advance\dimen6 .5\dimen0
        \setbox1=\vbox to\proofdotseparation{\vss\hbox{$\cdot$}\vss}%
        \setbox0=\hbox{%
                \advance\dimen6-.5\wd1
                \kern\dimen6
                $\vcenter to\proofdotnumber\proofdotseparation
                        {\leaders\box1\vfill}$%
                \unhbox\proofrulename}%
\else   \dimen6=\fontdimen22\the\textfont2 % height of maths axis
        \dimen7=\dimen6
        \advance\dimen6by.5\proofrulebreadth
        \advance\dimen7by-.5\proofrulebreadth
        \setbox0=\hbox{%
                \kern\shortenproofleft
                \ifdoubleproof
                \then   \hbox to\dimen0{%
                        $\mathsurround0pt\mathord=\mkern-6mu%
                        \cleaders\hbox{$\mkern-2mu=\mkern-2mu$}\hfill
                        \mkern-6mu\mathord=$}%
                \else   \vrule height\dimen6 depth-\dimen7 width\dimen0
                \fi
                \unhbox\proofrulename}%
        \ht0=\dimen6 \dp0=-\dimen7
\fi
\let\doll\relax
\ifwasinsideprooftree
\then   \let\VBOX\vbox
\else   \ifmmode\else$\let\doll=$\fi
        \let\VBOX\vcenter
\fi
\VBOX   {\baselineskip\proofrulebaseline \lineskip.2ex
        \expandafter\lineskiplimit\ifproofdots0ex\else-0.6ex\fi
        \hbox   spread\dimen5   {\hfi\unhbox\proofabove\hfi}%
        \hbox{\box0}%
        \hbox   {\kern\dimen2 \box\proofbelow}}\doll%
\global\dimen2=\dimen2
\global\dimen3=\dimen3
\egroup % NESTED ZERO
\ifonleftofproofrule
\then   \shortenproofleft=\dimen2
\fi
\shortenproofright=\dimen3
\onleftofproofrulefalse
\ifinsideprooftree
\then   \hskip.5em plus 1fil \penalty2
\fi
}

\usepackage{hyperref}

\newcommand{\CCC}{{\cal C}}
\newcommand{\DDD}{{\cal D}}

\newcommand{\RRR}{{\cal R}}
\newcommand{\SSS}{{\cal S}}

\newcommand{\XXX}{{\cal X}}
\newcommand{\YYY}{{\cal Y}}
\newcommand{\ZZZ}{{\cal Z}}
\renewcommand{\Bbb}{\mathbb}

\newcommand{\RRr}{{\Bbb R}}

\newcommand{\mathbold}[1]{\mbox{\boldmath $#1$}}

\mathcode`\<="4268 %left delimiter
\mathcode`\>="5269 %right delimiter
\mathchardef\gt="313E %relation >
\mathchardef\lt="313C %relation <
\newsavebox{\barr}
\savebox{\barr}{\hspace*{-9.5pt}\raisebox{1.25pt}{$
\scriptscriptstyle%
|$}\hspace*{4.5pt}} 
\newsavebox{\barrleft}
\savebox{\barrleft}{\hspace*{-8.5pt}\raisebox{1.25pt}{$
\scriptscriptstyle%
|$}\hspace*{10pt}}

 \def\pushright#1{{%              set up
    \parfillskip=0pt            % so \par doesnt push \square to left
    \widowpenalty=10000         % so we dont break the page before \square
    \displaywidowpenalty=10000  % ditto
    \finalhyphendemerits=0      % TeXbook exercise 14.32
    \leavevmode                 % \nobreak means lines not pages
    \unskip                     % remove previous space or glue
    \nobreak                    % don't break lines
    \hfil                       % ragged right if we spill over
    \penalty50                  % discouragement to do so
    \hskip.2em                  % ensure some space
    \null                       % anchor following \hfill
    \hfill                      % push \square to right
    {#1}                        % the end-of-proof mark (or whatever)
    \par}}                      % build paragraph

 \def\qed{\pushright{$\square$}\penalty-700 \smallskip}

\newenvironment{prf}[1]{\begin{trivlist} \item[{\bf ~Proof}#1.]}%
{\qed\end{trivlist}}

\newcommand{\be}[1]{\begin{#1}}

\newcommand{\ee}[1]{\end{#1}}
\newcommand{\beq}{\begin{equation}}
\newcommand{\eeq}{\end{equation}}
\newcommand{\ba}[1]{\begin{array}{#1}}
\newcommand{\ea}{\end{array}}
\newcommand{\bea}{\begin{eqnarray}}
\newcommand{\eea}{\end{eqnarray}}
\newcommand{\bear}{\begin{eqnarray*}}
\newcommand{\eear}{\end{eqnarray*}}
\newcommand{\bpr}{\begin{prf}{}}
\newcommand{\epr}{\end{prf}}
\newcommand{\bprf}[1]{\begin{prf}{#1}}
\newcommand{\eprf}{\end{prf}}

\theoremstyle{plain}
\newtheorem{thm}{Theorem}[section]
\newtheorem{prop}[thm]{Proposition}
\newtheorem{lemm}[thm]{Lemma}
\newtheorem{corr}[thm]{Corollary}

\newtheorem{defn}[thm]{Definition}
\theoremstyle{remark}

\renewcommand{\to}{\longrightarrow}

\newcommand{\tto}[1]{\xrightarrow{#1}}

\newcommand{\inclusion}{\!
\begin{tikzar}[column sep=1em]
\arrow[hookrightarrow]{r}\&\,\end{tikzar}\!\!}

\newcommand{\pfn}{\!
\begin{tikzar}[column sep=1em]
\arrow[rightharpoonup]{r}\&\,\end{tikzar}\!\!}%\newcommand{\pfn}{\rightharpoonup}
\newcommand{\ppfn}[1]{
\begin{tikzar}{}
\arrow[rightharpoonup]{r}{#1}\&\,\end{tikzar}}

\pgfdeclaredecoration{single line}{initial}{
        \state{initial}[width=\pgfdecoratedpathlength-1sp]{\pgfpathmoveto{\pgfpointorigin}}
        \state{final}{\pgfpathlineto{\pgfpointorigin}}
}

\tikzset{
        raise line/.style={
                decoration={single line, raise=#1}, decorate
        }
} 

\newcommand{\tores}{\, \begin{tikzpicture}[>=latex,->]
\draw[raise line=2,->] (0,0) -- +(.65,0);
\end{tikzpicture}\, }
\newcommand{\toress}{\, \begin{tikzpicture}[>=latex,->]
\draw[raise line=2,->] (0,0) -- +(1,0);
\end{tikzpicture}\, }

\renewcommand{\paragraph}[1]{\vspace{.3\baselineskip}\noindent\textbf{#1}}

\newcommand{\state}[2]{\left[{#1},{#2}\right]}

\newcommand{\undefined}[1]{{#1}\!\!\uparrow}

\newcommand{\Nod}{\SSS}
\newcommand{\Rou}{\XXX}
\newcommand{\Sou}{\YYY}
\newcommand{\ssou}{\beta}
\newcommand{\Zou}{\ZZZ}
\newcommand{\Var}{\Rou}%{\III}
\newcommand{\Val}{\Sou}%\OOO}

\newcommand{\alice}{A}
\newcommand{\bob}{B}
\newcommand{\carol}{C}
\newcommand{\dave}{D}
\newcommand{\experian}{E}
\newcommand{\gogol}{G}
\newcommand{\zuck}{Z}
\newcommand{\tizer}{T}
\newcommand{\subj}{S}

\newcommand{\Dist}{\Delta}
\newcommand{\Dis}{\DDD}

\newcommand{\ineq}[1]{\left\lfloor{#1}\right\rfloor}
\newcommand{\oueq}[1]{\left\lceil {#1}\right\rceil}
\newcommand{\reso}[1]{\llbracket{#1}\rrbracket}%{\varrho_{#1}}
\newcommand{\resoo}[2]{\reso{#1}^{(#2)}}
\newcommand{\req}{{\mathbold r}}%{\varrho_{#1}}
\newcommand{\poli}{{\mathbold p}}
\newcommand{\ssig}{\vartheta}
\newcommand{\SSig}[1]{\Theta(#1)}
\newcommand{\size}[1]{\#{#1}}
\newcommand{\ssize}[1]{\#\#{#1}}
\newcommand{\ssup}[1]{{#1}^{\#}}
\newcommand{\weight}[1]{\Sigma{#1}}

\newcommand{\djoin}{\curlyvee}%{\curlyveedownarrow}
\newcommand{\dmeet}{\curlywedge}%{\curlywedgeuparrow}
\newcommand{\Djoin}{\bigcurlyvee}%{\curlyveedownarrow}
\newcommand{\Dmeet}{\bigcurlywedge}%{\curlywedgeuparrow}
\newcommand{\fusion}{\widehat{\curlyvee}}%{\between}
\newcommand{\Fusion}{\widehat{\bigcurlyvee}}%{\between}
\newcommand{\jnt}{\raisebox{.25ex}{$\int$}}%{\curlyveedownarrow}
\newcommand{\pref}[1]{\stackrel{#1}\unlhd}%\triangleleft}
\newcommand{\prref}[1]{\stackrel{#1}\lhd}
\newcommand{\indif}[1]{\stackrel{#1}{\mbox{\begin{sideways}$\Diamond$\end{sideways}}}}

\begin{document}

\title{Privacy protocols}
\author{Jason Castiglione\thanks{Supported by NSF.}  \and Dusko~Pavlovic\thanks{Partially supported by NSF and AFOSR.} \and Peter-Michael Seidel\\
%	\email
	{\small {\rm Email:}~\{jcastig, dusko, pseidel\}@hawaii.edu} \\
	% \institute{
	University of Hawaii, Honolulu HI, USA}

\date{}

\maketitle

\begin{abstract} \noindent 
Security protocols enable secure communication over insecure channels. Privacy protocols enable private interactions over secure channels. Security protocols set up secure channels using cryptographic primitives. Privacy protocols set up private channels using secure channels. But just like some security protocols can be broken without breaking the underlying cryptography, some privacy protocols can be broken without breaking the underlying security. Such privacy attacks have been used to leverage e-commerce against targeted advertising from the outset; but their depth and scope became apparent only with the overwhelming advent of influence campaigns in politics. The blurred boundaries between privacy protocols and privacy attacks present a new challenge for protocol analysis. Covert channels turn out to be concealed not only below overt channels, but also above: subversions, and the \emph{level-below}\/ attacks are supplemented by sublimations and the \emph{level-above}\/ attacks. 
\end{abstract}

\section{Introduction: What is privacy?}\label{Sec:Intro}
% !TEX root = 0-PrivT.tex

The concept of privacy has been a source of much controversy and confusion, not only in social and political discourse, but also in research. 

The \emph{controversy}\/ arises from the fact that privacy is not a security requirement, like secrecy or authenticity, but a fundamental \emph{social right}. In the US jurisprudence, justices Warren and Brandeis~\cite{Warren-Brandeis} defined it as the \emph{right to be left alone}. Privacy is thus not just a technical task of controlling some assets, but first of all a \emph{political}\/ task, requiring that some policies should be specified and implemented. Privacy policies are generally designed to balance public and private interests, assets, and resources. This balancing sometimes comes down to playing out the political forces against one another. %This is why the history of mankind is the history of wars, often arising from the never shifting boundaries and demarkation lines between the public and the private sphere. 

The \emph{confusion}\/ around privacy also arises from the fact that it is both a technical and a political problem. The problems of privacy are discussed in many different research communities, often in different terms, or with the same terms denoting different concepts. Privacy provides an opportunity for security researchers to contribute to political discourse~\cite{RogawayP:moral}. It also provides politicians an incentive to get involved in the technical discourse about security. The overarching source of confusion are the political narratives constructed the process of shifting and blurring the boundaries between the public and the private, which has been the driving force behind social transformations for centuries%of social revolutions that begin by declaring that the private sphere does not exist, and of dictatorships that totally privatize the public sphere
~\cite{ArendtH:human,HabermasJ:thesis}.

\paragraph{History of privacy.}
Social history is first of all the history of shifting demarcation lines between the public  sphere and the private sphere~\cite{BaileyJ:private,orlin2009locating,schoeman1984philosophical}. Communist revolutions usually start by abolishing not only private property, but also private rights. Tyrannies and oligarchies, on the other hand, erode public rights and ownership, and privatize  resources and social life. The distinction between the realm of public (city, market, warfare\ldots) and the realm of private (family, household, childbirth\ldots) 
%the realms of 
%\begin{itemize}
%\item public: city, market, warfare\ldots and
%\item private: family, household, childbirth\ldots 
%\end{itemize}
was established and discussed in antiquity \cite{angela2009rome,burke2000delos}. It was a frequent topic in Greek tragedies: e.g. Sophocles' \emph{Antigone}\/ is torn between her private commitment to her brothers and her public duty to the king. The English word \emph{politics}\/ comes from the Greek word \textgreek{p'olis}, denoting the public sphere; the English word \emph{economy}\/ comes from the Greek word \textgreek{o'ikos}, denoting the private sphere.

\paragraph{Distinguishing privacy.} There are many aspects of privacy, conceptualized in different research communities, and studied by different methods; and perhaps even more aspects that are not conceptualized in research, but arise in practice, and in informal discourse. We carve out a small part of the concept, and attempt to model it formally.

As an abstract requirement, privacy is a negative constraint, in the form \emph{"bad things should not happen"}. Note that secrecy and confidentiality are also such negative constraints, whereas authenticity and integrity are positive constraints, in the form \emph{"good things should happen"}. More precisely, authenticity and integrity require that some desirable information flows happen. E.g., a message \emph{"I am Alice"}\/ is authentic if it originates from Alice. On the other hand, confidentiality, secrecy and privacy require that some undesirable information flows are prevented: Alice's password should be secret, her address should be confidential, and her health record should be private. 

But what is the difference between privacy, secrecy and confidentiality? Let us first move out of the way the difference between the latter two. In the present paper, we ignore that difference. In the colloquial usage, the terms confidentiality and secrecy allow subtle distinctions: e.g., when a report is confidential, we don't know its contents; but when it is secret, we don't even know that it exists. 
% But this is a paper about privacy as a right. 
Secrecy and confidentiality are both security properties. We bundle them into one and use them interchangeably. (Restricting to just one of them gets awkward.)

As for the difference between privacy and secrecy (or confidentiality), it comes in two flavors:
\begin{description}
\item[1)] while secrecy is a property of \emph{information}, privacy is a property of any \emph{asset}\/ or \emph{resource}\/ that can be secured\footnote{Information is, of course, a resource, so it can be private.}; and
\item[2)] while secrecy is a \emph{local}\/ requirement, usually imposed on information flowing through a given channel, privacy is a \emph{global}\/ requirement, usually imposed on all resource requests and provisions, along any channel of a given network. 
\end{description}
Let us have a look at some examples.

\paragraph{Ad (1)}, Alice's password is secret, whereas her bank account is private. Bob's health record is private, and it remains private after he shares some of it with Alice. It consists of his health information, but it may also contain some of his tissue samples for later analysis. On the other hand, Bob's criminal record is in principle not private, as criminal records often need to be shared, to protect the public. Bob may try to keep his criminal record secret, but even if he succeeds, it will not become private. Any resource can be made private if the access to it can be secured. E.g.,  we speak of a private water well, private funds, a private party if the public access is restricted. On the other hand, when we speak of a secret water well, secret funds, or a secret party, we mean that the public does not have any information about them. A water well can be secret but not private, or it can be private, but not secret; or it can be both, or neither.

\paragraph{Ad (2)}, to attack Alice's secret password, Bob eavesdrops at the channel where she uses it; to attack her bank account, he can, of course, impersonate Alice using her password; but he can also initiate a request through any of the channels of the banking network that he can access; best of all, he can coordinate an attack through many channels. To attack a secret, a cryptanalyst analyzes a given cipher. To attack privacy, a data analyst gathers and analyzes data from as many surveillance points as he can access. To protect secrecy, the cryptographer must assure that the plaintexts cannot be derived from the ciphertexts without the key, for a given cipher. To protect privacy, the network operator must assure that there are no covert channels anywhere in the network. 
%
%Privacy is only preserved if all network channels are secure. Privacy is the requirement that some private rights are preserved everywhere in a given network. While secrecy is a locally testable property, predicated over the inputs and the outputs of a given channel, privacy is not locally testable, as it can fail, e.g. if private data can be compiled from many localities in a network, but none of them discloses it alone. 
%
%
%
%But large open networks, like the web, or the internet, spread beyond anyone's horizon of observation. We are living in a network of such networks. %There are always more channels beyond the horizon. So how do we dam remote flows? 
%What does it mean to protect privacy?

\paragraph{Defining privacy.} Secrecy is formally defined in cryptography. The earliest definition, due to Shannon~\cite{ShannonC:Secrecy}, says that it is a property of a channel where the outputs are statistically independent of the inputs. It is tempting, and seems natural, to define privacy in a similar way. This was proposed by Dalenius back in the 1970s~\cite{DaleniusT:desideratum}. A database is private if the public data that it discloses publicly say nothing about the private data that it does not disclose. This \emph{desideratum}, as Dalenius called it, persisted in research for a number of years, before it became clear that it was generally impossible as a requirement. E.g., if everybody knows that Alice eats a lot of chocolate, but there is an anonymized database that shows a statistical correlation between eating a lot of chocolate and heart attacks, then this database discloses that Alice may be at a risk of heart attack, which should be Alice's private information, and thus breaches Dalenius' desideratum. Notably, this database breaches Alice's privacy \emph{even if}\/ Alice's record does not come about in it. Indeed, it is not necessary that Alice occurs in the database either for establishing the correlation between chocolate and heart attack, or for the public knowledge that Alice eats lots of chocolate; the two pieces of information can arise independently. Alice's privacy can be breached by linking two completely independent pieces of information, one about Alice and chocolate, the other one about chocolate and heart attacks. But since Alice's record does not come about in the database, it cannot be removed from it, or anonymized in it. Making sure that neither Alice nor any other record can be identified closer than up to a set of $k$ other records with the same attributes, as required by the popular $k$-anonymity approach to privacy~\cite{SweeneyL:kanonymity,SweeneyL:suppression}, would not make any difference for Alice's privacy in this case either, since Alice cannot be identified at all in a database where she does not come about. Since Alice's privacy is not breached by identifying her, but by linking her public attribute (chocolate) with the public statistic correlating that attribute with a private attribute (heart disease), it follows that anonymity cannot assure privacy.
%
%The first difference between secrecy (or confidentiality) and privacy is that the former is a property of data flowing through a channel, whereas the latter is the property of whatever may be flowing through the whole network, which consists of many channels. The essence can a data network, or a market, as a network transporting money and goods; or an energy network, or a traffic network. Data privacy is only preserved if the data remain confidential in all channels of the network. Privacy is the requirement that some private rights are preserved everywhere in a given network. While secrecy is a locally testable property, predicated over the inputs and the outputs of a given channel, privacy is not locally testable, as it can fail, e.g. if private data can be compiled from many localities in a network, but none of them discloses it alone. To attack a secret, a cryptanalyst analyzes a given cipher. To attack privacy, a data analyst can gather and analyze data from many surveillance points. To protect secrecy, a cryptographer must assure that plaintexts cannot be derived from ciphertexts without the key. To protect privacy, a network operator must assure that there are no covert channels in the network, local, or nonlocal. But large open networks, like the web, or the internet, spread beyond anyone's horizon of observation. We are living in a network of such networks. %There are always more channels beyond the horizon. So how do we dam remote flows? 
%What does it mean to protect privacy?

\subsubsection*{Notation.}
It is convenient to view each natural number as the set of the numbers preceding it, i.e. $n= \{0,1,2,\ldots, n-1\}$.

\section{Resources and their fusion}
\label{Sec:resources}

\subsection{Concepts}
\be{defn}
A \emph{source element}\/ of a set $\Sou$ is a function $\ssou:\Sou \to [0,1]$ such that 
\begin{itemize}
\item the set $\ssup\ssou = \{y\in \Sou\ |\ \ssou(y) \neq 0\}$ is finite, and 
\item the sum $\weight \ssou = \sum_{y\in \Sou} \ssou(y)$ is not greater than 1. 
\end{itemize}
The set of all source elements of $\Sou$ is denoted by $\Dis \Sou$.  The set $\ssup\ssou$ is called the \emph{support} of $\ssou$, and its number of elements $\size\ssou$ is called the \emph{size} of $\ssou$. The number $\weight\ssou$ is the \emph{total weight} of $\ssou$.  The values $\ssou(y)$ are \emph{weights}\/ (or \emph{probabilities}) of $y\in \Sou$.
\ee{defn}

Ordinary elements $y\in \Sou$ correspond to the source elements $\chi_y :\Sou\to [0,1]$ where $\chi_y(y) = 1$ and $\chi_y(z) = 0$ for all $z\neq y$.  A source element $\ssou$ that happens to be \emph{total}, in the sense that $\sum_{y\in \Sou} \ssou(y) = 1$, corresponds to what would normally be called a finitely supported probability distribution over $\Sou$. The set of all total source elements of $\Sou$ is $\Dist \Sou$. They are often viewed geometrically, as points of the convex polytope spanned by  $y \in \Sou$ as  vertices. If we adjoin to $\Sou$ a fresh element $\ast$, and thus form the set $\Sou' = \Sou \cup \{\ast\}$, then each source element $\ssou$ of $\Sou$ (not necessarily total) can be mapped into a total source element $\ssou'$ of $\Sou'$, defined
\bear
\ssou'(y) & = & \begin{cases}
\ssou(y) & \mbox{ if } y\in \Sou\\
1-\weight\ssou & \mbox{ if } y = \ast
\end{cases}
\eear
It is easy to see that this gives a bijection
\bear
\Dis \Sou & \cong & \Dist \Sou'
\eear
so that the source elements of  $\Sou$ can be viewed as the points of the convex polytope over $\Sou'$. Extended along this bijection, the inclusion $\Dist \Sou \subseteq \Dis \Sou$ becomes the retraction 
$\Dist\Sou \, \raisebox{-1pt}{$\rightarrowtail$}\hspace{-1.1em}\raisebox{2pt}{$\twoheadleftarrow$}\, \Dist\Sou'$, which projects the polytope $\Dist\Sou'$ to the face where the weight of $\ast$ is 0.

This geometric view of source elements of $\Sou$, as convex combinations from $\Sou'$, supports the intuition that they are \emph{'incomplete'}\/ probability distributions, which don't add up to 1. The probability deficiency $1-\Sigma\ssou$ can be thought of as the chance that sampling the source element $\ssou$ does not yield any output.

\be{defn}\label{Def:resource}
A\/ \emph{resource} is a function $\varphi :\Rou\to \Dis \Sou$ whose \emph{support} $\ssup \varphi = \{x\in \Rou\ |\ \varphi_x \neq 0\}$ is finite. The number of elements of $\ssup \varphi$ is called the \emph{size} of $\varphi$ again, and its \emph{total size} is the number 
\bear \ssize \varphi &= & \sum_{x\in \ssup\varphi} \size{\varphi_x}\eear
\ee{defn} 

A resource that takes as its values the ordinary elements, i.e. the source elements which take a value 1, is just a partial function from  $\Rou$ to $\Sou$. A resource that takes only total source elements as its values can be viewed as a stochastic matrix, i.e. a matrix of numbers from the interval $[0,1]$ where columns add up to 1. Any resource $\varphi :\Rou\to \Dis \Sou$ can be viewed as a $\Rou\times \Sou$-matrix of numbers $\varphi_x(y) \in [0,1]$. Although the sets $\Rou$ and $\Sou$ can be infinite, the finiteness of $\size \varphi$ and of all $\size{\varphi_x}$ implies that only $\ssize \varphi$ many entries of this infinite matrix are different from 0. Any given resources $\varphi :\Rou\to \Dis \Sou$ and $\psi: \Sou\to \Dis \Zou$ can thus be composed into a resource $\psi\varphi : \Rou \to \Dis \Zou$, obtained by matrix composition
\bear
(\psi\varphi)_x(z) & = & \sum_{y\in \ssup\psi} \varphi_x(y) \cdot \psi_y(z)
\eear
\paragraph{Notation.} Since we will most of the time look at the functions in the form $\Rou\to \Dis \Sou$, let us omit the $\Dis$, and write such functions as $\Rou \tores \Sou$, denoting by the arrow with full head $\tores$ a function whose outputs are source elements. Such functions are our resources.

\subsection{Examples}
While the notion of resource is very general, we begin with some very special cases, to ground the intuitions.

\paragraph{Example 0:} Simple communication channels are the most basic examples of resources. More precisely, a \emph{memoryless channel}, as defined in any information theory textbook, is a resource $\varphi: \Rou\tores \Sou$, where $\Rou$ and $\Sou$ are finite alphabets, and all source elements $\varphi_x$ are total. E.g., the binary symmetric channel is such a resource over the alphabets $\Rou = \Sou = \{0,1\}$, and with the elements distributed by 
\bear
\varphi_x(y)\ =\ \Pr\left(y|x\right) & = & \begin{cases} 
1- p & \mbox{ if } x= y\\
p & \mbox{ otherwise }
\end{cases}
\eear
A channel that depends on state, but with no feedback, can also be viewed as a resource, at least at each finite step, since its outputs always depend on finite histories, and are thus finitely supported. Asynchronous channels can be modeled using source elements that may not be total. However, our path in this paper leads in a different direction.

\paragraph{Example 1:}
A \textbf{database} is a function $\varphi : \RRR \times \CCC\tores \Val$, where
\begin{itemize}
\item $\RRR$ is a set of \emph{rows}, or \emph{records}, or \emph{items},
\item $\CCC$ is a set of \emph{columns}, or \emph{attributes}, 
\item $\Val$ is a set of \emph{values}\/ or  \emph{outputs}.
\end{itemize}
The "row-column" terminology suggests that we think of a database as an $\RRR\times \CCC$-matrix of entries from $\Dis\Val$. When the attributes $c\in \CCC$ are expressed using different sets of values $\Val_c$, we take  $\Val = \bigcup_{c\in \CCC} \Val_c$.  For ordinary databases, the entries are ordinary elements $\varphi_{rc}\in \Val$, or they may be empty, and the resource $\varphi$ is an ordinary map, or a partial map. For online databases, it can be uncertain what will be returned in response to a query, since the data are updated dynamically, often concurrently; so the entries are stochastic, and we model them as source elements $\varphi_{rc}\in \Dis\Val$.  

In fact, the whole web can be viewed as a large, dynamic, stochastic database.

\paragraph{Example 2:}
The \textbf{web} is a resource $\omega : \Var \tores \Val$ where
\begin{itemize}
\item $\Var$ is the set of URLs (or more precisely of all possible HTTP requests),
\item $\Val$ is the set of HTML documents (embedded in the HTTP responses, often extended with JavaScript executables, JSON or XML object references, etc.) 
\end{itemize}
The web is, of course, a very complex, very dynamic environment, and when you input a URL $x\in \Var$ into your browser, the process that determines what output will the random variable $\omega_x$ return to your browser is ongoing nonlocally, and there is a lot of uncertainty and chaos in it, like in any complex natural process, such as the weather. The web is a typical source showing why privacy is so hard: it combines politics and thermodynamics.

\paragraph{Example 3:}
A \textbf{search engine} $\gamma : \Var \tores \Val$ (e.g. Google) is an attempt at  a map of the web:
\begin{itemize}
\item $\Var$ is the index of keywords and search terms built from them,
\item $\Val$ is the set of indexed web pages.
\end{itemize}
The distribution of $\gamma_x\in \Dis\Val$ over the set of the hits for the search term $x\in \Var$ captures the web page \emph{ranking} \cite{page98pagerank,PavlovicD:CSR08}: higher ranked pages are assigned higher probabilities in $\gamma_x$. 

\paragraph{Example 4:}
A \textbf{social engine} $\zeta : \Var \tores \Val$ (e.g. Facebook) is a shared resource for social networking through posting messages, media, and gestures, and distributing them according to some specified privacy policies along the social channels provided by the platform. In the most basic model,
\begin{itemize}
\item $\Var$ is the set of identifiers of all users' postings, 
\item $\Val$ are the contents of the postings, i.e. the set of the posted messages, media, and gestures.
\end{itemize}
For simplicity, we assume that the identifiers $x\in \Var$ contain all needed references to their sources, and that the contents of the postings are either equal (i.e., they can be repeated) or different, but have no intrinsic structure or correlations. A social engine can thus be viewed as an ordinary mapping $\zeta : \Var \to \Val$ of identifiers as ordinary elements $x\in \Var$ into the postings as ordinary elements $\zeta_x\in \Val$. The randomness will emerge from sharing: who sees whose postings. And sharing is determined by running privacy protocols.

\subsection{Resource fusion}\label{sec:fusion}
Data sources are naturally ordered by the amount of information that they provide: e.g., a database $\varphi$ may provide Alice's IP address, say 98.151.86.153; a database $\psi$ may provide just the first block: 98. The amount of information is usually quantified by the entropy of its source: e.g., Alice's record $\varphi_\alice$ would contain  32 bits of information, whereas Alice's record $\psi_\alice$ would contain only 8 bits. However, if the IP prefix given by $\psi_\alice$ is different from the IP prefix given by $\varphi_\alice$, and if we are interested in the actual address, then counting the bits does not suffice. Data analysts do not just \emph{quantify}\/ the amounts of information in the available data, they also \emph{qualify}\/ them, and compare their information contents. To model that practice, we need an ordering $\prec$ where $\psi_\alice\prec \varphi_\alice$ only if the information provided by $\psi_\alice$ is really contained in $\varphi_\alice$, not just more uncertain; and moreover, we need an operation $\psi_\alice\fusion \varphi_\alice$ for \emph{information fusion}, which will join together the parts of $\psi_\alice$ and $\varphi_\alice$ where they are consistent with each other, and discard the parts where they contradict each other. This may sound like a tall order in theory, but that is what data analysts do in practice. 

The problems of ordering information sources turn out to have been studied, albeit implicitly, in the theory of \emph{majorization}  \cite{Alberti-Uhlmann:book,Olkin:book}. The basic techniques go back to \cite{hardy-littlewood-polya}, where a host of inequalities from different parts of mathematics were derived using majorization, as if by magic. In the meantime, it has been well understood that the power of magic was due to ordering information sources, but the problem of conjoining and reconciling information sources has not yet been been directly addressed, although some technical results and conceptual expositions came close to it \cite{Ando:majoriz,NielsenM:char-mixing-majoriz}.
%\raggedbottom
%
%\vfil\penalty-20\vfilneg
\subsubsection{Preferences and consistency}
\be{defn}\label{def:preference}
Every source element $\ssou\in \Dis\Sou$ induces the following binary relations:
\begin{itemize}
\item \emph{strict preference}: $u\prref\ssou v \ \iff \  \ssou(u) \lt \ssou(v)$;
\item \emph{preference}: $u\pref\ssou v \ \iff \  \ssou(u) \leq \ssou(v)$;
\item \emph{indifference}: $u\indif\ssou v \ \iff \  \ssou(u) = \ssou(v)$.
\end{itemize}
\ee{defn}
Clearly, $\prref\ssou$ is transitive, $\pref\ssou$ is also reflexive; and $\indif\ssou$ is moreover symmetric, and thus an equivalence relation. The preference relation $\pref\ssou$ is thus a preorder on $\Sou$ and a partial order on the set of $\indif\ssou$-equivalence classes $\Sou/\indif\ssou$. For $U,V\subseteq \Sou$ we thus write
\bear
U\prref \ssou V &\iff & \forall u u' \in U\ \forall v' v\in V.\ \ u\indif \ssou u'\wedge u'\prref\ssou v' \wedge v' \indif\ssou v
\eear
and ditto for $U\pref \ssou V$.
% We will find useful the ability to compare sources based upon the above orderings. To illustrate the comparison we will extend strict preference to equivalence classes under indifference.  Given a source $\ssou$, we can display the partial order on equivalence classes  as \[ \{ u | \ssou(u)=\alpha_1 \} \prref\ssou  \{u | \ssou(u) = \alpha_2 \} \prref\ssou \dots \prref\ssou \{  u | \ssou(u) = \alpha_k  \} .\]  The associated sets are equivalence classes under indifference, and \[ \{ \alpha_1 ,\alpha_2 , \dots ,\alpha_k | 0 \lt \alpha_1 \lt \alpha_{2} \lt \dots \lt \alpha_k  \} = \{ \ssou(u) | u \in  \ssup\ssou \} .\] Without loss of generality, we will not include the values for which the source is zero, i.e., $ \Sou  \setminus \ssup\ssou$.

\be{defn}\label{def:consistency}
For a finite set of source elements $B\subset \Dis\Sou$, we define
\bear
u \pref B v &  \iff & u\pref{\ssou_0}w_1 \pref{\ssou_1}w_2\pref{\ssou_2}\cdots w_n\pref{\ssou_n} v\\
&& \hspace{2em}\mbox{ for some } \ssou_0, \ssou_1,\ldots, \ssou_n\in B\mbox{ and }w_1, \ldots, w_n\in \Sou\\ 
u \indif B v &  \iff & u \pref B v \mbox{ and } v \pref B u 
\eear
Then we say that the set $B$ is \emph{consistent} if
\bear
u \prref{\ssou} v \mbox{ for some } \ssou \in B & \implies & u \pref{\delta} v \mbox{ for all } \delta \in B
\eear
In addition, we define a $\lhd$-\emph{cycle} in $B$  as $u_0 \prref{\delta_1} u_1 \prref{\delta_2} \dots u_{n-1} \prref{\delta_n} u_{n}$ where $u_i \in \Sou$ and  $\delta_i \in B$.
\ee{defn}

\paragraph{Comment.}  Inconsistency means that taking the transitive closure of the relations $\pref\ssou$ for $\ssou\in B$ creates new $\lhd$-cycles of preferences. For instance, if $B = \{\ssou,\delta\}$ are such that  
\begin{itemize}
\item $a\prref \ssou b$ and $c\prref \ssou d$, but 
\item $b\prref\delta c$ and $d\prref \delta a$, 
\end{itemize}
then $\pref B$ contains the $\lhd$-cycle $a\prref\ssou b\prref\delta c\prref\ssou d \prref\delta a$. This $\lhd$-cycle makes $a, b, c$ and $d$ all $\indif{\ssou,\delta}$-equivalent, while it is easy to assign the weights in $\ssou$ and  $\delta$ in such a way that no pair of elements is either $\indif\ssou$-equivalent or $\indif\delta$-equivalent.

\paragraph{Example.}
Let  $\ssou, \gamma, \delta \in \Dis\Sou$ be sources with support $\{w,x,y,z \} \subset \Sou$. Let their values on the support be defined in the table below.

\[\begin{array}{c|llll}
	& w & x & y  & z   \\
	\hline  \ssou & 0.1 & 0.2 & 0.2 & 0.3  \\
	\gamma & 0.1 & 0.1 & 0.3 & 0.5 \\ 
	\delta & 0.6 & 0.1 & 0.1 & 0.2
\end{array} \]

To illustrate consistency, we display the strict preference order on equivalence classes below.
\[\begin{array}{ccccc}
 \{ w \} & \prref\ssou & \{x,y\}  &  \prref\ssou & \{z\}   \\
\{ w ,x \} & \prref\gamma & \{ y\}  &  \prref\gamma & \{z\}   \\
\{ x,y \} & \prref\delta & \{z \}  &  \prref\delta & \{w \}  
\end{array} \]
 
 Let $B= \{ \ssou , \gamma \}$, then the transitive closure results in the total order  
 \[\{ w,x,y \}  \prref B  \{z\} \]
 Observe $z  \prref\delta w $, yet $w  \prref\ssou z $ and $w  \prref\gamma z $. Thus $\delta$ is not consistent with neither $\ssou$ nor $\gamma$, and therefore not $B$.

%\raggedbottom
%
%\vfil\penalty-20\vfilneg
\be{defn}\label{def:ordering}
An\/ \emph{ordering} $\ssig$ of a source element $\ssou:\Sou \to [0,1]$ is a pair of functions $
%\beq\label{eq:CCC}
\begin{tikzar}{}%[row sep=4em,column sep=4em]
N \arrow[bend right = 12,tail]{r}[swap]{\ssig}\& \Sou  \arrow[bend right = 12,two heads]{l}[swap]{\widetilde\ssig} 
\end{tikzar}
%\eeq
$ such that 
\begin{itemize}
\item $N\gt \size\ssou$;
\item for all $i\in N = \{0,1,\ldots,N-1\}$ holds $\widetilde \ssig \ssig(i) = i$ ;
\item for all $y\in \ssup \ssou$ holds $\ssig\widetilde \ssig(y) = y$,
\item $\ssou{\ssig(0)} \geq \ssou{\ssig(1)} \geq  \cdots \geq \ssou{\ssig(\size\ssou-1)}\gt \ssou{\ssig(\size\ssou)}=\cdots = \ssou{\ssig(N-1)}= 0$.
\end{itemize}
The set of all orderings of $\ssou \in \Dis \Sou$ is denoted by $\SSig\ssou$.
\ee{defn} 

We hope that this formal definition does not conceal the idea of ordering a source element, which is as simple as it sounds: an ordering $\ssig$ of  $\ssou:\Sou\to [0,1]$ enumerates the support $\ssup\ssou$ by the indices from the set
%\footnote{Recall that we view each number as the set of its predecessors, and thus $N = \{0,1,2,\ldots, N-1\}$.}
  $N = \{0,1,\ldots,\size\ssou-1, \size \ssou,\ldots,N-1\}$ in such a way that $u\in \ssup\ssou$ are ordered by weight, with those with the highest weights $\ssou(u)$ coming first. All $u\in \Sou$ for which $\ssou(u)=0$, and thus  $u\not\in \ssup\ssou$, are mapped to $\ssig(u)\geq \size\ssou$.  Any indifferent pair $u,v \in \Sou$, i.e. such that $\ssou(u) = \ssou(v)$ and thus $u \indif\ssou v$, allows 2 different orderings. Any indifference class with $k$ elements allows $k!$ different orderings. On the other hand, a source element $\ssou$ where for any pair $u\neq v\in \ssup \ssou$ holds $\ssou(u)\neq \ssou(v)$ induces a unique ordering of  its support $\ssup\ssou$. Indeed,  it is easy to show that in that case, all orderings $\ssig,\xi\in \SSig\ssou$ must satisfy $\ssou{\ssig(i)} = \ssou{\xi(i)}$ for all $i\lt \size \ssou$, and thus induce the same ordered sequence $\ssou\ssig = \ssou \xi$. This unique ordered version of $\ssou$ is usually written $\ssou^\downarrow$, but we will write $\ssou\ssig$ when an explicit ordering is needed, as it will be the case in Prop.~\ref{prop:cons-order}. Identifying the set $\ssup \ssou \subseteq \Sou$ with the set of numbers $\size \ssou = \{0,1,\ldots,\size\ssou-1\}\subset N$ allows writing the source element $\ssou:\Sou\to [0,1]$ as the descending sequence
\bear
\ssou^\downarrow \ =\  \ssou\ssig\  & = &\  \big< \ssou{\ssig(0)},\ssou{\ssig(1)}, \ldots , \ssou{\ssig(\size \ssou-1)}, 0, 0, 0,\ldots \big>
\eear
Note that "un-ordering" $\ssou\ssig$ by $\widetilde\ssig$ restores $\ssou$, because for all $y\in \ssup\ssou$ holds $\ssou{\ssig\widetilde\ssig(y)} = \ssou(y)$. We can also consider $\ssou^\downarrow $ as a source over the natural numbers, $\ssou^\downarrow: \mathbb{N} \rightarrow [0,1]$. Given a set $B \subset \Dis\Sou$ and  the unique ordered version of the source elements, we may construct a set $B^\downarrow=\{\ssou^\downarrow | \ssou \in B\} \subset \Dis\mathbb{N}$. Observe, given any finite set of source elements the above set will always have a common ordering. Let $N \gt \size B $. Then an example of a common ordering is $\ssig(i)=i$ for all $i \in N $, $\widetilde\ssig (i)=i$ for $i \lt N$, and $\widetilde\ssig (i)=N-1$ for $i \gt N$.

\paragraph{Example.} Let $ \Sou $ be the set of URLs and $\ssou:\Sou \to [0,1]$, a source element. Let $\ssou$ represent the probability of visiting a URL first for a certain user when they open a web browser. 

\[\begin{array}{c|l}
\text{URL} &\ssou  \\
\hline 
 \text{http://www.google.com} & 1/4 \\
 \text{https://www.amazon.com} & 1/4\\ 
\text{https://en.wikipedia.org} & 1/4 \\
\text{https://stackoverflow.com/}& 1/8
\end{array} \]

Let $\ssou$ be zero for all other URLs. Observe that the probabilities do not add up to one. The intent is to capture what happens if they immediately close the browser, or say the computer crashes and the user is unable to visit a URL. Given the above ordering we may define   $\ssig(0),\ssig(1),\ssig(2) ,\ssig(3), \ssig(4)$ accordingly. Thus we have $\ssou^\downarrow \ =\ \big<  \frac{1}{4} , \frac{1}{4} ,\frac{1}{4} ,\frac{1}{8} ,0\big>$. The resulting total order is \[\{  \ssig(3) \} \prref\beta \{\ssig(0),\ssig(1),\ssig(2) \} \]

The orderings of $\ssou$ that are extended to $N\gt \ssup\ssou$ just add more indices than there are in  $\size\ssou$, and thus cover by the enumeration $\ssig$ not only the support $\ssup\ssou$, but also some $y\in \Sou$ for which $\ssou(y) = 0$. Such extensions are needed when we look for common orderings of $\Sou$ induced by different source elements $\ssou, \gamma :\Sou \to [0,1]$. Two such source elements may have different supports $\ssup\ssou$ and $\ssup \gamma$, but their orderings may be consistent, in the sense that they both may extend to the  same ordering of $\ssup\ssou \cup \ssup\gamma$.  Such an extended ordering $\ssig$ with $N$ enumerating $\ssup\ssou \cup \ssup\gamma$ would belong both to $\SSig\ssou$ and to $ \SSig\gamma$, and would thus be an element of the intersection $\SSig\ssou \cap \SSig\gamma$. It will turn out that this intersection characterizes the situation when $\ssou$ and $\gamma$ are consistent, and can be conjoined into a single source: see Prop.~\ref{prop:cons-order} below.

%\raggedbottom
%
%\vfil\penalty-5\vfilneg
\subsubsection{Majorization preorder}
% \hspace{2em}\\[2ex]
\paragraph{Sequence differentials and integrals.} To define joins and meets of source elements, we borrow the following operations from \cite{PavlovicD:LICS98}. Let $\RRr^\ast$ be the set of sequences of reals. The operations 
$\jnt, \partial : \RRr^*\to \RRr^*$ map any sequence $\big<\alpha(0),\alpha(1),\ldots, \alpha(n-1)\big>$ to its \emph{integral version}  $\big<\jnt\alpha(0),\ldots, \jnt\alpha({n-1})\big>$ and its \emph{differential version} $\big<\partial\alpha(0),\ldots, \partial\alpha({n-1})\big>$ defined
\[
\jnt\alpha(k) \ = \ \sum_{i=0}^{k} \alpha({i})\qquad\qquad\qquad
\partial\alpha(k) \ = \ \alpha(k) - \alpha(k-1)
\]
for $0\leq k\lt n$. We assume $\alpha({-1}) = 0$. Note that $\jnt\partial(\alpha) = \alpha = \partial\jnt(\alpha)$.

\begin{prop}\label{thm:majorization}
For any $\ssou, \gamma \in \Dis \Sou$, and $n = \max(\size\ssou,\size\gamma)$  the following conditions are equivalent:
\begin{enumerate}[label=\alph*),ref=(\alph*)]
\item $\ssou = D\gamma$, where $D$ is a doubly substochastic matrix\footnote{A $\Sou\times \Sou$-matrix with finitely many nonzero, nonnegative entries is doubly stochastic if the sums of the entries in each nonzero row and in each nonzero column are 1. Already Garrett Birkhoff considered infinite doubly stochastic matrices, asking for the infinitary generalization of his doubly stochastic decomposition in the problem 111 of  his \emph{Lattice Theory}.};\label{stat1}
\item $\ssou = \sum_{i=0}^{n-1} \lambda_i P_i \gamma$ where  $\lambda_i\in [0,1]$, $\sum_{i=0}^{n-1} \lambda_i \leq1$, and $P_i$ are partial permutations, i.e., a submatrix of a permutation matrix;\label{stat2}
\item $\jnt\ssou^\downarrow{(k)} \leq \jnt \gamma^\downarrow{(k)}$ for all $k\lt n$.\label{stat3}
\end{enumerate}
\end{prop}

%\bpr
%Observe, given a doubly substochastic matrix, $D$, we may form a doubly stochastic matrix, $D'$, as follows. Given a $n \times 1$ vector, $x$, let $\text{diag}(x)$ denote the matrix with $x_i$'s on the diagonal and zeros elsewhere. Let $\bar{1}$ denote the all $1$'s vector. Then we may define a doubly stochastic matrix.  \[D'= \begin{bmatrix} 
%D & I - \text{diag}(D \bar{1}) \\
%I- \text{diag}( D^\intercal \bar{1})& D^\intercal
%\end{bmatrix} \]
%
%The equivalence of conditions \ref{stat1} and  \ref{stat2} follows from applying Birkhoff's Theorem \cite{birkmaj} to $D'$, which states that the set of doubly stochastic matrices is the convex hull of the permutation matrices. The equivalence of conditions  \ref{stat1} and \ref{stat3} follows from Hardy, et al., \cite{hlp:ineq, hlp:simpleconvex}. The essence of their proof decomposes $D'$ as a product of transformations. Each transformation takes two points, say $u$ and $v$, where $\gamma(u) \lt \gamma(v)$. Then the transformation scales $\gamma$ to $\gamma'$ such that $\gamma'(v)-\gamma'(u) \lt \gamma(v)-\gamma(u)$ but leaves $\gamma$ unchanged elsewhere. They proceed to show that transforming $\ssou$ by a finite sequence of these transformations results in $\beta$ if and only if $\beta$ is majorized by $\ssou$, i.e., condition \ref{stat3}. 
%\epr

\begin{defn}\label{def:majoriz}
When source elements $\ssou$ and $\gamma$ satisfy any of the equivalent conditions of Thm.~\ref{thm:majorization}, we say that $\ssou$ is \emph{majorized}\footnote{This is also referred to as weak majorization, see for example \cite{marshallmajor}.} by $\gamma$ and write $\ssou \prec \gamma$.
%
%
%The \emph{majorization} of source elements is the binary relation $\prec$ on   
%
%
%satis 
%\bea\label{eq:majoriz-def}
%\ssou \prec \gamma\ & \iff &\ \exists \ssig \in \SSig\ssou \cap \SSig\gamma . \   
% \forall k \lt \size\ssou + \size\gamma.\ \ \sum_{i=0}^k \ssou{\ssig(i)} \leq \sum_{i=0}^k \gamma{\ssig(i)}
%\eea
\ee{defn}
%\begin{lemm} If $\ssou \prec \gamma$, then there exists a source element, $\zeta$, with the %same shape as $\ssou$ such that $\ssou \prec \zeta \prec \gamma$. Furthermore it is maximal %in the sense, that any other $\zeta'$ satisfying the same criteria has $\zeta' \prec \zeta$.
%\end{lemm}
%\paragraph{Proof Idea.}

%
%
%\paragraph{Source ordering.} For simplicity, first consider source elements over finite sets. Since a finite set can always be ordered, we assume that each set with $n$ elements given in the form $n = \{0,1, \ldots, n-1\}$. 
%
%\begin{lemma} Given $\gamma \in \Dis n$, let $\widehat \gamma$ be its ordered version, i.e.
%\begin{itemize}
%\item $\{\gamma_0,\ldots, \gamma_{n-1}\} = \{\widehat\gamma_0,\ldots, \widehat\gamma_{n-1}\}$
%\item $\widehat \gamma_0 \geq \widehat \gamma_1\geq\cdots \widehat\gamma_{n-1}$
%\end{itemize}
%The for all $k\in [1, n]$ holds 
%\bear \sum_{i=0}^{k-1} \widehat \gamma_i & = &   \bigvee_{t_0,\ldots, t_{n-1} \in [0,1]} \left\{\sum_{i=0}^n t_i \gamma_i\  |\ \sum_{i=0}^n t_i = k\right\}
%\eear
%\end{lemma}
%
%
%
%Given source elements $\beta, \gamma \in \Dis n$, define
%\bea
%\beta \prec \gamma & \iff & \forall \ell \lt n. S^\ell \beta \leq S^\ell \gamma
%\eea
%where $S^\ell \beta = \sum_{i=1}^\ell \beta_i$.

%\raggedbottom
%
%\vfil\penalty-10\vfilneg
\subsubsection{Meets, joins, and fusions}

%\paragraph{Consistent preference preorders have  common refinements.} 
%It is clear from the definition that a set of source elements $B$ is consistent if and only if the equivalence classes with respect to $\indif{B}$ are obtained as the intersections of the equivalence classes with respect to $\indif{\ssou}$ for $\ssou\in B$. 

Since all $\pref{\ssou}$ are \emph{total} preorders, in the sense that any two elements of $\Sou$ are comparable, and since each of them thus induces a strict total order on its indifference equivalence classes, it follows that the total preorder $\indif{B}$ is the least common refinement of all partitions of $\Sou$ induced by $\ssou \in B$, in the sense that it induces a strict total order on its own of equivalence classes, which are the least common refinement of $\indif{\ssou}$ for $\ssou\in  B$. The task is now to lift the least common refinement $\pref B$ of the preference preorders $\pref\ssou$ induced by source elements $\ssou \in B\subset \Dis \Sou$ to a least lower bound $\djoin B$ with respect to a suitable information preorder $\prec$. 

\begin{prop}\label{prop:join-meet}
	Let $B\subset \Dis\Sou$ be a finite set of source elements such that $\bigcap\limits_{\beta \in B} \SSig\beta$ is not empty. Let  $\begin{tikzar}{}%[row sep=4em,column sep=4em]
	M \arrow[bend right = 12,tail]{r}[swap]{\ssig}\& \Sou  \arrow[bend right = 12,two heads]{l}[swap]{\widetilde\ssig} 
	\end{tikzar}$ be any one of the common orderings in the intersection.  The meet and join of $B$ with respect to the majorization preorder are respectively
	\[
	\Dmeet B \ = \ \Big(\partial \bigwedge_{\ssou \in B} \jnt\ssou\ssig \Big){\widetilde\ssig}
	\qquad\qquad\qquad\qquad \Djoin B \ = \ \Big(\partial\bigvee_{\ssou\in B} \jnt\ssou\ssig \Big){\widetilde\ssig}
	\]
	where $\bigwedge$ and $\bigvee$ are pointwise.
\end{prop}

\begin{corr}\label{cor:all-join-meet}
Let $B\subset \Dis\Sou$ be a finite set of source elements.  There exists at least one common ordering $\begin{tikzar}{}%[row sep=4em,column sep=4em]
M \arrow[bend right = 12,tail]{r}[swap]{\ssig}\& \Sou  \arrow[bend right = 12,two heads]{l}[swap]{\widetilde\ssig} 
\end{tikzar}$ in $\bigcap\limits_{\ssou^\downarrow \in B^\downarrow} \SSig{\beta^\downarrow}$. Their meet and join with respect to the majorization preorder are respectively
\[
\Dmeet B^\downarrow \ = \ \partial \bigwedge_{\ssou^\downarrow \in B^\downarrow} \jnt\ssou^\downarrow 
\qquad\qquad\qquad\qquad \Djoin B^\downarrow \ = \ 
\partial\bigvee_{\ssou^\downarrow   \in B^\downarrow} \jnt\ssou^\downarrow 
\]
where $\bigwedge$ and $\bigvee$ are pointwise and $\Dmeet B^\downarrow , \Djoin B^\downarrow  \in \Dis \mathbb{N}$
\end{corr}

Corollary \ref{cor:all-join-meet} states that given any finite set of source elements, the meet and join always exist for the unique order. Furthermore, this meet and join will help quantify the inconsistency of a finite set of source elements.

\paragraph{Relating majorization to fusion.} The fact that majorization is a reflexive and transitive relation, i.e. a \emph{preorder}\/ on $\Dis\Sou$, is just a slight refinement of the classical results of \cite{hardy-littlewood-polya} and \cite{BirkhoffG:doubly}. In the meantime, the role of majorization as \emph{information}\/ preorder has also been well established in several research communities \cite{NielsenM:char-mixing-majoriz}. Our main interest here is to understand and model the operation of \emph{fusion}\/ of resources, used in the practices of data analysis, and in the attacks on private resources. The usual scenario is that an analyst acquires two or more resources, and fuses them together, in order to extract as much information or value as possible. Intuitively, this corresponds to finding a least upper bound (usually called \emph{supremum}, or \emph{join}) with respect to the majorization preorder of the acquired resources. But a pair of sources may not have any upper bounds with respect to majorization, and therefore cannot always be conjoined together. The following proposition characterizes that situation as inconsistency. Prop.~\ref{prop:imposing-consistency}, coming right after, describes what can be done to make a set of sources consistent.

\begin{prop}\label{prop:cons-order}
	The following conditions on a finite set $B\subset \Dis\Sou$ are equivalent:
	\begin{enumerate}[label=\alph*),ref=(\alph*)]
		\item $B$ is consistent (in the sense of Def.~\ref{def:consistency});
		\item there is a common ordering $\ssig \in \bigcap_{\ssou\in B}\SSig\ssou$ of $B$ (in the sense of
		Def.~\ref{def:ordering});
		\item $\exists$ $\theta: \mathbb{N} \rightarrow \Sou $, such that, $\jnt\Dmeet B^\downarrow  = \bigwedge\limits_{\ssou \in B}\jnt\ssou \theta$;
		\item $\exists$ $\theta: \mathbb{N} \rightarrow \Sou $, such that, $ \jnt\Djoin B^\downarrow  = \bigvee\limits_{\ssou \in B}\jnt\ssou \theta$.
	\end{enumerate}
	%\end{alpherate}
\end{prop}

Given a finite set of sources, what can we do if they are inconsistent? Suppose we are predicting a hurricane path, and several models each predict fundamentally different paths. How do we extract the information that is consistent between all models? Given $B \subset \Dis\Sou$ is finite, our first step is to partition the source $\Sou$, so that we may find the common information on the partition elements. Each set in the partition is defined in a minimal way so that sources in $B$ must be made constant on them to fuse the consistent information. Given $y \in \Sou$ we define the \emph{consistency class} of $B$ at $y$ as the set
\[B_y = \{y\} \cup \{ u \in \Sou \ | \ u \text{ is in a} \lhd\text{-cycle with }y \text{ in }B \}  \]

\begin{lemm}\label{lem:cclass}
	Given $B \subset \Dis\Sou$ is finite, the consistency classes of $B$, $\{B_y \}_{y \in \Sou}$, are a  partition of $\Sou$. Furthermore, if $B_u \cap B_v = \emptyset$, then either $B_u \pref{\delta}B_v$ or  $B_v \pref{\delta} B_u$ for all $\delta \in B$.

\end{lemm}

The following proposition captures how to make a set of sources consistent while not increasing the individual probabilities of any element. Furthermore, the fixed points will correspond to consistent sets.

\begin{prop}\label{prop:imposing-consistency}
	Let $B \subset \Dis\Sou$ be a finite set. Given $\ssou \in B$ define the function $\widehat \ssou : \Sou \to [0,1]$ by
	\bear
	\widehat \ssou(y) & = & \bigwedge_{  u \in B_y }  \ssou(u)
	\eear
Then $\widehat \ssou$ is the greatest among all source elements $\alpha \prec \ssou$ which are consistent with the elements of $B$. Moreover, the set $\widehat B = \left\{\widehat \ssou\ |\ \ssou \in B\right\}$ is consistent. 
	
\end{prop}

%\paragraph{Proof} First, we show $\widehat \ssou$ is consistent with each element of $B$. If $u \prref{\widehat\ssou} v $, then $B_u \neq B_v$.

% $u \prref{B} v$. Hence $u \prref{\delta} v$ for any $\delta \in B$. On the other hand, suppose $u \prref{\delta} v$ for some $\delta \in B$. If $u \indif{B}v$ then $u \indif{\widehat\ssou}v$. Otherwise, then we must have since 

%It is clear by the definition that $\widehat \ssou (v) \leq \ssou(v)$ for all $v \in \Sou$ and therefore is majorized by $\ssou$. This is important since it preserves the probabilities that remain. Now we show that $\widehat\ssou$ is the greatest among all source elements majorized by $\ssou$, consistent with $B$ and satisfying the above inequality. 

%Suppose $\alpha$ satisfies the above and is bounded element-wise by $\ssou$. If $ \alpha (v) \leq \widehat\ssou(v)$ for all $v \in \Sou$, then we are done. Let us assume there exists $v \in \Sou$ such that $ \widehat\ssou(v) \lt \alpha (v) \leq \ssou(v)$. By definition of $\widehat\ssou$ we have that $u \indif{B} v$ for some $u \in U$, since otherwise $ \widehat\ssou(v) = \ssou(v)$. Hence  $\alpha (u) \leq \ssou(u)= \widehat\ssou(v) \lt \alpha(v) \leq \ssou(v)$. Then observe either $v \in U$ or $v \notin U$. If $v \in U$, then we immediately have a contradiction since there exists $\delta \in B$ such that $v \prref{\delta} u$. This implies that $\alpha$ is inconsistent with $\delta$ which is a contradiction. Thus $\widehat\ssou(v) = \alpha (v)$ for all $v \in \Sou$.

\begin{corr}
	 Any finite set of source elements $B\subset \Dis\Sou$ has a majorization meet $\Dmeet B \ = \ \Dmeet \widehat B$, whereas its meet is constructed as in Prop.~\ref{prop:join-meet}.
\end{corr}
\begin{defn}\label{def:source-fusion}
The\/ \emph{source fusion} of a finite set $B\subset \Dis\Sou$ is the source element $\Fusion B\in \Dis\Sou$ defined by
\bear
\Fusion B & = & \Djoin \widehat B 
\eear
where $\widehat B$ is the consistent set constructed as in Prop.~\ref{prop:imposing-consistency}.
\end{defn}

\noindent A concrete and familiar \textbf{example} of this operation will be described in Sec.~\ref{sec:alibi}.  

\be{defn}\label{def:resource-fusion}
The \emph{resource fusion}\/ $\Fusion \Phi :\Rou \tores \Sou$ of  a finite set of resources $\Phi \subset \Rou \tores \Sou$  is defined pointwise along $x\in \Rou$ by
\bear
\left(\Fusion \Phi\right)_x & = & \Fusion \Phi_x 
\eear
where the source fusion on the right-hand side is from Def.~\ref{def:source-fusion}. 
\end{defn}

%\begin{defn}
%Let $\Phi\subset \Rou \tores \Sou$ be a finite set of resources. Its fusion is the resource $\fusion \Phi :\Rou \tores \Sou$ defined at $x\in \Rou$ by
%\bear
%\fusion \Phi_x & = & 
%\Djoin_{\varphi \in \Phi} \widehat{\varphi_x} 
%\eear
%\end{defn}

\paragraph{Inconsistent sources generate a higher-order source.} Another view of the above definition is that  the set of resources $\Phi$ is the pointwise join of the set of resources $\widehat \Phi$, where for each $x\in \Rou$, the set of sources $\widehat \Phi_x$ is the greatest consistent set  under $\Phi_x$. The notion of consistency of sources, as imposed in Def.~\ref{def:consistency} requires a consensus of all sources, wherever they declare their preferences by assigning different weights. This is a very restrictive notion of consistency. Different domains require different notions of fusion. Through centuries, many different forms of preference aggregation have been proposed and are nowadays systematized and analyzed in theories of voting and social choice \cite{saari1995basic,suzumura2009rational,brandt2016handbook}. An important aspect not analyzed within that tradition are the \emph{higher-order}\/ source aggregations. While with the preference aggregations and the source fusion operations like the one presented above, inconsistent sources are conjoined into a less informative source, any inconsistencies observed in hypothesis testing are the source of new information within a higher-order source \cite{popper2002conjectures}. See Sec.~\ref{sec:alibi} for an example.

 The only point of even mentioning this vast conceptual area in this constrained space is in support of our claim that the rapidly evolving practices of information gathering and analysis require a theory of information that takes into account the information content and quality, and not just the transmission rate measures, and quantity. %There is much more to be desired than increasing the channel capacity.

\section{Privacy protocols for private channels}
\label{Sec:protocols}
% !TEX root = 0-PrivT.tex

\subsection{Concepts}
Interactions, communication, and resource sharing are usually modeled using networks. A \emph{network}\/ structure is based on a graph, here consisting of 
\begin{itemize}
\item a set $\Nod = \{\alice, \bob, \carol,\dave,\ldots, \subj,\ldots\}$  of \emph{nodes}, representing subjects or users, often called Alice, Bob, Carol, Dave\ldots; and
\item a set of links $\alice \tto{} \bob$, representing the channels between Alice and Bob.
\end{itemize}
In most examples, there will be at most one channel between any pair of subjects, so the network structure boils down to a binary relation on the set $\Nod$. In any case, channels allow us to model \emph{local}\/ interactions: Alice can interact with Bob only if there is a channel $\alice \tto{} \bob$, in which case we say that Bob is Alice's \emph{neighbor}. A network is often assumed to provide routing services, whereby Alice can send a message or object to Dave, who is not her neighbor, from neighbor to neighbor: $\alice \tto{} \bob \tto{}\carol \tto{} \dave$. The Internet is, of course, a network with routing services. 

A network can be viewed as infrastructure for sharing some private resources. The node Alice has a private resource $\reso\alice:\Var_\alice \tores \Val_\alice$, but she can achieve more if she cooperates with Bob, who has a resource  $\reso\bob:\Var_\bob \tores \Val_\bob$, and can share some of it with Alice, often in exchange for some  of hers. Privacy protocols are abstract models of such transactions. Optimizing utility of their private resources, all rational agents engage in such transactions at all network levels.  The obvious examples are market transactions, where Alice and Bob trade goods, money, labor; also health care, insurance, rescue missions, regulatory control, search, education. At lower levels of interaction, many basic social phenomena arise from privacy protocols. But in a data-driven society, certain data privacy protocols take a life of their own, while other privacy protocols become invisible, and private. %Like security, privacy grows into a principal-agent problem

The basic building block of private interactions is a 2-message protocol pattern, depicted in Fig.~\ref{Fig:privproto}, where Alice submits to Bob a \emph{request} $\req^{\alice\bob}$ for some of his private resource $\reso\bob$, and Bob responds according to his \emph{policy} $\poli^{\alice\bob}$, and shares with Alice the part of his resource that results from composing her request and his policy, i.e. $\reso{\alice\bob} = \poli^{\alice\bob}\reso\bob\req^{\alice\bob}$. Our claim is that all privacy protocols are obtained by composing suitable instances of this pattern. The idea is that these \emph{Request-Policy (RP)}-components are atoms of privacy, just like the \emph{Challenge-Response (CR)}-components are atoms of authentication  \cite{PavlovicD:CSFW04,PavlovicD:CSFW05,PavlovicD:ISTPS08,PavlovicD:CWSP09,PavlovicD:ICDCIT12}. The incremental approach to protocol design, analysis, taxonomy, and to security proofs \cite{PavlovicD:JCS05,PavlovicD:JCS04,PavlovicD:ESORICS04} seems to extend naturally from authentication and key establishment protocols to privacy protocols. 

\subsection{Basic privacy protocols}
Bob's private resource $\reso\bob:\Var_\bob \tores \Val_\bob$ accepts Bob's private inputs from $\Var_\bob$ and produces Bob's private outputs in $\Val_\bob$. In order to be able to request access  to some of Bob's private resource, Alice must be able to  reference some of his private inputs, and to  observe and utilize some of his private outputs. To allow subjects to refer to each other's private resources, we distinguish
\begin{itemize}
\item sets $\Var$ and $\Val$ of \emph{global}\/ identifiers; and
\item sets $\Var_\subj$ and $\Val_\subj$ of \emph{local}\/ identifiers, owned by each $\subj\in \Nod$, and accessed by the owner using the functions
\beq\label{eq:privproj}
\begin{tikzar}{}%[row sep=4em,column sep=4em]
\Var_\subj \arrow[bend right = 12,tail]{r}[swap]{\underline\pi_\subj}\& \Var  \arrow[bend right = 12,two heads,dashed%,rightharpoondown
]{l}[swap]{\overline\pi_\subj} 
\end{tikzar}
\qquad \qquad \qquad\qquad\qquad
\begin{tikzar}{}%[row sep=4em,column sep=4em]
\Val_\subj \arrow[bend right = 12,tail]{r}[swap]{\underline\rho_\subj}\& \Val  \arrow[bend right = 12,two heads,dashed%rightharpoondown
]{l}[swap]{\overline\rho_\subj} 
\end{tikzar}
\eeq
such that $\overline \pi_\subj \underline\pi_\subj(x) = x$ and $\overline \rho_\subj \underline\rho_\subj(y) = y$.\end{itemize}
The last two equations make $\underline \pi_\subj$ and $\underline\rho_\subj$ total, but $\overline \pi_\subj$ and $\overline\rho_\subj$ can be genuinely partial, in which case the composites $\pi_\subj = \underline\pi_\subj\overline \pi_\subj$ and $\rho_\subj = \underline \rho_\subj \overline\rho_\subj$ are also partial. It is easy to see that they are also idempotent, i.e. satisfy $\pi_\subj\pi_\subj=\pi_\subj$ and $\rho_\subj\rho_\subj=\rho_\subj$.
%, and moreover 
%\beq\label{eq:privid}
%\Var_\subj = \{x\in \Var\ |\ \pi_\subj(x) = x\} 
%\qquad \qquad \quad
%\Val_\subj = \{y\in \Val\ |\ \rho_\subj(y) = y\} \eeq
The following definition puts together all of the above.
%
%\be{defn}
%A \/ \emph{projector} is a partial function $\pi:X\pfn X$ which is idempotent, i.e. $\pi\circ \pi = \pi$.
%\ee{defn} 

\be{defn}\label{def:network}
A\/ \emph{resource network} consists of 
\begin{itemize}
\item a \emph{network} $\Nod = \{\alice, \bob, \carol,\ldots, \subj,\ldots\}$;
\item \emph{global identifiers} $\Var = \{x,x',\ldots\}$ for inputs and $\Val = \{y,y',\ldots\}$ for outputs;
%\item a set $\Var$ of \emph{public input identifiers},
%\item a set $\Val$ of \emph{public output identifiers},
\item for each network node $\subj\in \Nod$,  the local castings like in \eqref{eq:privproj}, which induce
\begin{itemize}
\item the projector $\pi_\subj: \Var\pfn\Var$ determining the \emph{local input identifiers} $\Var_\subj = \{x\in \Var\ |\ \pi_\subj(x) = x\}$, and
\item  the projector $\rho_\subj: \Val\pfn\Val$ determining the \emph{local output identifiers} $\Val_\subj  = \{y\in \Val\ |\ \rho_\subj(y) = y\}$, 
\item a\/ \emph{resource} $\reso \subj: \Var \tores \Val$. 
\end{itemize}
\end{itemize}
\ee{defn}

\paragraph{Remark.} Alice's resource $\reso \alice$ provides her with the output $\rho_\alice\reso \alice  x$ on input $x\in \Var_\alice$, and in many cases it satisfies $\reso\subj = \rho_\subj \reso\subj \pi_\subj$, or equivalently $\rho_\subj \reso\subj  = \reso\subj  = \reso\subj \pi_\subj$. However, it may happen that $x,x'\in \Var$ are indistinguishable for Alice as inputs, i.e. $\pi_\alice(x) = \pi_\alice(x')$, but that they give her different outputs i.e. $\rho_\alice\reso \alice ( x)\neq \rho_\alice\reso \alice (x')$. Intuitively, this means that the global identifiers may \emph{interfere}\/ with the environment in ways that are for Alice not directly observable, but that she may be able to observe such interferences indirectly, from the outputs of her resource. This will be discussed in Sec.~\ref{Sec:nonint}.

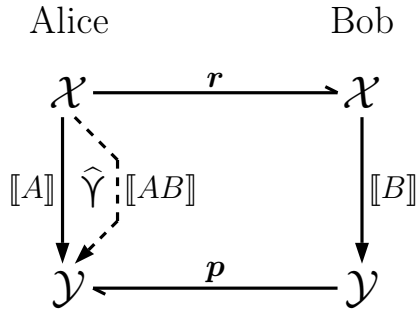
\begin{figure}[h!t]
\centering
\newcommand{\Alce}{\large Alice}
\newcommand{\Vars}{{\Large$\Var$}}
\newcommand{\Vals}{{\Large$\Val$}}
\newcommand{\Bobo}{\large Bob}
\newcommand{\Alceresource}{\reso{\alice}}
\newcommand{\Alcebobsresource}{\reso{\alice\bob}}
\newcommand{\Bobsresource}{\reso{\bob}}
\newcommand{\Request}{\req}
\newcommand{\Response}{\poli}
\newcommand{\addjoin}{\Fusion}
\def\JPicScale{.65}
\ifx\JPicScale\undefined\def\JPicScale{1}\fi
\psset{unit=\JPicScale mm}
\psset{linewidth=0.3,dotsep=1,hatchwidth=0.3,hatchsep=1.5,shadowsize=1,dimen=middle}
\psset{dotsize=0.7 2.5,dotscale=1 1,fillcolor=black}
\psset{arrowsize=1 2,arrowlength=1,arrowinset=0.25,tbarsize=0.7 5,bracketlength=0.15,rbracketlength=0.15}
\begin{pspicture}(0,0)(76.25,60)
\psline[linewidth=0.65](20,45)(70,45)
\psline[linewidth=0.65](20,5)(70,5)
\psline[linewidth=0.65,arrowsize=1.55 2,arrowlength=1.5,arrowinset=0]{->}(75,40)(75,10)
\rput(15,60){\Alce}
\rput(75,60){\Bobo}
\rput[b](45,46.25){$\Request$}
\rput[b](45,6.25){$\Response$}
\rput[l](76.25,25){$\Bobsresource$}
\psline[linewidth=0.65,linestyle=dashed,dash=2 2](25,31.25)(25,18.75)
\rput[r](12.5,25){$\Alceresource$}
\psline[linewidth=0.65,linestyle=dashed,dash=2 2](16.25,40)(25,31.25)
\psline[linewidth=0.65,linestyle=dashed,dash=2 2,arrowsize=1.5 2,arrowlength=1.5,arrowinset=0.05]{->}(25,18.75)(16.25,10)
\psline[linewidth=0.65,arrowsize=1.55 2,arrowlength=1.5,arrowinset=0]{->}(13.75,40)(13.75,10)
\psline[linewidth=0.65](67.5,46.25)(70,45)
\psline[linewidth=0.65](20,5)(22.5,3.75)
\rput[l](26.25,25){$\Alcebobsresource$}
\rput(75,45){\Vars}
\rput(15,5){\Vals}
\rput(75,5){\Vals}
\rput(15,45){\Vars}
\rput(20,25.62){$\addjoin$}
\end{pspicture}
\caption{Basic privacy protocol: Request-Policy (RP)}
\label{Fig:privproto}
\end{figure}

%
%
%
%
%
%The idea is that the set of identifiers is a disjoint union $\Var = \coprod_{\subj\in \Nod} \Var_\subj$ of the inputs $\Var_\subj$ available at all network nodes. It is thus assumed that the names of all Bob's identifiers from $\Var_\bob \subset \Var$ are available to Alice, but not their values in $\Val$, since Alice's private resource $\reso\alice : \Var \to \Val$ is only defined on Alice's private identifiers $\Var_\alice \subset \Var$, and undefined for elsewhere. Knowing the names of Bob's identifiers allows Alice to request their values from Bob in a privacy protocol. The set $\Var$ thus plays the role of a "global API", or an "abstract DNS".\footnote{This global name space can be construed as the privacy version of the Kerckhoffs' principle: \emph{no privacy by obscurity}.} The idea is that, when Alice requests and obtains some of Bob's private resources, this is captured in the model by expanding the definition domain of her resource map $\reso\alice$, to include the value assignments to some of Bob's identifiers from $\Var_\bob\subset \Var$. This is modeled in the next section.
%

\be{defn}
A \/\emph{Request-Policy (RP)} protocol between Alice and Bob in a resource network $\Nod$ is a pair of partial functions $\Phi = \left<\req_\Phi^{\alice\bob}, \poli_\Phi^{\alice\bob}\right>$, where
\begin{itemize}
\item $\req_\Phi^{\alice\bob}: \Var\pfn \Var$ is Alice's \emph{request} for Bob's private resource, and
\item $\poli_\Phi^{\alice\bob}: \Val \pfn \Val$ is Bob's privacy \/\emph{policy}\/ towards Alice.
\end{itemize}
The\/ \emph{outcome} of a run of the privacy protocol is that the part of Bob's private resource that is approved by % in the domain of 
Bob's policy and referenced %in the range of
by Alice's request is released to her, providing her with the resource:
\bea\label{eq:AB}
\reso{\alice\bob} & = &
\big(\Var \ppfn{\req_\Phi^{\alice\bob}} \Var \stackrel{\reso{\bob}}{\toress} \Val \ppfn{\poli_\Phi^{\alice\bob}} \Val\big)
\eea
which is then conjoined with Alice's own resource $\Var\stackrel{\reso{\alice}}{\toress} \Val$, thus providing Alice with the total resource $\reso\alice \Fusion \reso{\alice\bob}$,
as displayed in Fig.~\ref{Fig:privproto}. 
\ee{defn}

\begin{figure}[h!t]
\centering
\newcommand{\Vars}{{\large$\Var$}}
\newcommand{\Vals}{{\large$\Val$}}
\newcommand{\Request}{\req}
\newcommand{\Response}{\poli}
\newcommand{\downwards}{\mbox{\LARGE$\Downarrow$}}
\newcommand{\upwards}{\mbox{\LARGE$\Uparrow$}}
\def\JPicScale{.6}
\vspace{.5\baselineskip}
\ifx\JPicScale\undefined\def\JPicScale{1}\fi
\psset{unit=\JPicScale mm}
\psset{linewidth=0.3,dotsep=1,hatchwidth=0.3,hatchsep=1.5,shadowsize=1,dimen=middle}
\psset{dotsize=0.7 2.5,dotscale=1 1,fillcolor=black}
\psset{arrowsize=1 2,arrowlength=1,arrowinset=0.25,tbarsize=0.7 5,bracketlength=0.15,rbracketlength=0.15}
\begin{pspicture}(0,0)(165,41.25)
\psline[linewidth=0.65](10,40)(60,40)
\psline[linewidth=0.65](10,0)(60,0)
\psline[linewidth=0.65,arrowsize=1.55 2,arrowlength=1.5,arrowinset=0]{->}(65,35)(65,5)
\rput[t](35,38.75){$\Request$}
\rput[b](35,1.25){$\Response$}
\psline[linewidth=0.65,linestyle=dashed,dash=2 2](15,26.25)(15,13.75)
\psline[linewidth=0.65,linestyle=dashed,dash=2 2](6.25,35)(15,26.25)
\psline[linewidth=0.65,linestyle=dashed,dash=2 2,arrowsize=1.5 2,arrowlength=1.5,arrowinset=0.05]{->}(15,13.75)(6.25,5)
\psline[linewidth=0.65,arrowsize=1.55 2,arrowlength=1.5,arrowinset=0]{->}(3.75,35)(3.75,5)
\psline[linewidth=0.65](57.5,41.25)(60,40)
\psline[linewidth=0.65](10,0)(12.5,-1.25)
\rput(65,40){\Vars}
\rput(5,0){\Vals}
\rput(65,0){\Vals}
\rput(5,40){\Vars}
\rput(35,20){$\downwards$}
\psline[linewidth=0.65](110,40)(160,40)
\psline[linewidth=0.65](110,0)(160,0)
\psline[linewidth=0.65,arrowsize=1.55 2,arrowlength=1.5,arrowinset=0]{->}(165,35)(165,5)
\rput[t](135,38.75){$\Request$}
\rput[b](135,1.25){$\Response$}
\psline[linewidth=0.65,linestyle=dashed,dash=2 2](115,26.25)(115,13.75)
\psline[linewidth=0.65,linestyle=dashed,dash=2 2](106.25,35)(115,26.25)
\psline[linewidth=0.65,linestyle=dashed,dash=2 2,arrowsize=1.5 2,arrowlength=1.5,arrowinset=0.05]{->}(115,13.75)(106.25,5)
\psline[linewidth=0.65,arrowsize=1.55 2,arrowlength=1.5,arrowinset=0]{->}(103.75,35)(103.75,5)
\psline[linewidth=0.65](157.5,41.25)(160,40)
\psline[linewidth=0.65](110,0)(112.5,-1.25)
\rput(165,40){\Vars}
\rput(105,0){\Vals}
\rput(165,0){\Vals}
\rput(105,40){\Vars}
\rput(135,20){$\upwards$}
\end{pspicture}
\caption{Demand-driven protocols vs supply-driven protocols}
\label{Fig:forward}
\end{figure}
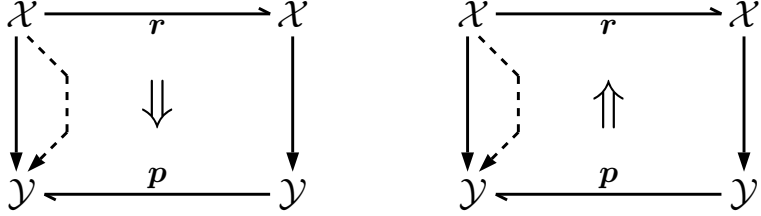

\paragraph{Privacy protocols can be driven by demand, or by supply.} The two interactions displayed in Fig.~\ref{Fig:privproto}, corresponding to the request $\req$ and the policy $\poli$ may happen in time in either order: some RP-protocols are initiated by the requester Alice, whereas other are initiated by the provider Bob. Fig.~\ref{Fig:forward} displays the two directions of in which the protocols can be executed as the hollow arrows.
E.g., a seller Bob may advertise his goods, provide samples, and influence the buyer Alice to request to buy more. In a demand-based economy, the buyers stroll the market place and request to buy the good that the sellers display, in a supply-based economy, the buyers are passive, and the sellers supply their goods, and try to create the demand, e.g. by presenting to the buyers previously unknown goods. But even in the traditional market place, neither side is passive, and the demand and supply are actively balanced to maximize profits on each side: the sellers look for buyers, the buyers shop around, they all haggle, they try to outsmart each other, which gives rise to deception.

\subsection{Examples of basic protocols}\label{sec:examples-prots}
\paragraph{Example 1:}
Let  $\varphi : \RRR \times \CCC\tores \Val$ be the database of a credit rating agency $\experian$xavier, presented as a resource in the following format:
\begin{itemize}
\item $\RRR = \Nod$, i.e. the database rows correspond to the network nodes;
\item $\CCC = \{T_0, T_1,\ldots, T_{n}\}$ are the types of financial and other relevant  transactions;
\item $\Var = \Nod \times \CCC$, i.e. the public identifiers $x\in \Var$ are in the form $x = <\alice, T_i>$, or $T_i^\alice$, denoting a type of Alice's transactions;
\item $\Var_\subj = \{T^\subj_0, T^\subj_1,\ldots, T^\subj_n\}$, i.e. $\pi_\subj(T_i^\alice) = T_i^\alice$ if $\subj = \alice$ or $\subj =\experian$, otherwise it is undefined, meaning that by $\reso\alice = \reso\alice \pi_\alice$ Alice can see her own record, and Exavier can see all records;
\item $\Val =\coprod_{i=0}^n T_i^\ast$ are the transaction history listings, unencrypted,  and
\item $\Val_\subj = \Val$, i.e. $\rho_\subj(y) = y$ for all $y\in \Val$ makes any released transaction listing readable to anyone.
%, where 
%\item $\Val_\subj = \coprod_{i=0}^n T_\subj^n \cup \{\ast\}$ is the type of history listings $t^i_\subj:T^i_\subj$ or the redacted listing $\ast$, so that the projection  $\rho_\bob\left<t_\alice, t_\bob,\ldots, t_\subj,\ldots \right> = \left<\ast, t_\bob, \ldots, \ast,\ldots\right>$ 
\end{itemize}
If Alice is a lender and Bob has requested a loan, then Alice may request access to some of his credit history by $\req^{\alice\bob}: \Var\pfn \Var$ with $\req^{\alice\bob}(x) = x$ if $x \in \Var_\bob$, otherwise undefined. Bob's privacy policy $\poli^{\alice\bob}: \Val\pfn \Val$ will then determine which transaction types $T_i$ will be released to Alice by  setting $\poli^{\alice\bob}(t) = t$ when $t\in T^\ast_i$ for some chosen values of $i$, otherwise undefined; or he may set his policy to release just some partial information about some of the transaction types. To prevent that Bob tampers with this information, this privacy protocol for obtaining loan applicants' credit rating information is in practice combined with an authentication protocol, where Bob's transaction record $t\in T^\ast_i$ is released to Alice not by Bob, but by the credit rating agency Exavier himself, upon Bob's approval. Exavier then uses the opportunity to record Alice's request in Bob's credit rating, and thus expands his shared database.

In any case, Alice the lender is provided the resource $\reso{\alice\bob} = \rho^{\alice\bob}\reso\bob \pi^{\alice\bob}$, controlled by Bob and Exavier. If Bob has joint accounts with Carol, Alice may request and Bob and Exavier may provide some information about Carol. Alice will then conjoin the obtained information with her own information and resource $\reso\alice$ and use the compiled information $\reso{\alice}\Fusion \reso{\alice\bob}$, or maybe $\reso{\alice}\Fusion \reso{\alice\bob}\Fusion\reso{\alice\bob\carol}$, to make her lending decision. Note that the described Request-Policy protocol is thus embedded within another RP-protocol, where Bob requests from Alice a loan, and Alice requests from Bob some private data, and uses the obtained information resources $\reso{\alice}\Fusion \reso{\alice\bob}\Fusion\reso{\alice\bob\carol}$ as the input into her loan policy, which outputs the loan decision. The privacy protocol for credit rating is thus composed with an authentication protocol for data release, and embedded into a privacy protocol for loan provision.

\paragraph{Example 2:}
The web as a resource $\omega : \Var \tores \Val$ is controlled by Bob through $\pi_\bob:\Var\pfn\Var$, which filters the set of URLs $\Var_\bob$ under Bob's control, locating his web site. The projector $\rho_\bob:\Val\pfn\Val$ determines what is published on his web site $\reso\bob = \rho_\bob\omega\pi_\bob$. If Alice surfs or navigates to Bob's web site, she submits a request $\req^{\alice\bob}:\Var\pfn\Var$ for some of Bob's URLs. Bob may then request some of her identifiers, and maybe some of her private data to authenticate her, and input the obtained information into his policy $\poli^{\alice\bob}:\Val\pfn\Val$ which generates and outputs Bob's web content $\reso{\alice\bob} = \poli^{\alice\bob}\reso\bob\req^{\alice\bob}$ in response to Alice's request. Alice can now add what she got from Bob to her own resource $\reso\alice = \rho_\alice\omega\pi_\alice$, and use $\reso\alice \Fusion \reso{\alice\bob}$ to serve instances of her own web site, when requested in other runs of the web protocol. 

\paragraph{Example 3:}
If Bob owns the search engine $\gamma : \Var \tores \Val$, then his private resource is simply $\reso\bob = \gamma$. If anyone can submit any search term from $\Var$, and if all pages indexed in $\Val$ are publicly accessible, then $\Var_\subj = \Var$ and $\Val_\subj = \Val$ for all subects $\subj$ on the network $\Nod$. A session of the RP-protocol thus consists of a query, represented as the partial function $\req^{\alice\bob}:\Var\pfn\Var$, undefined everywhere except on some search term $x = \req^{\alice\bob}(x)$; and a reply according to a policy $\poli^{\alice\bob}:\Val\pfn\Val$. In modern search engines, the search results are \emph{personalized}, in the sense that our search engine Bob tailors his response especially for Alice. This may mean that Bob's policy $\poli^{\alice\bob}$ is to display just what is of interest for Alice, or what she wants to hear. To achieve this, $\poli^{\alice\bob}$ skews the sample of the source element $\omega_x\in \Dis\Val$, output by the search engine $\omega:\Var\tores\Val$ in response to the query $x$; \emph{or}\/ it modifies the induced preference ordering $\prref{\omega_x}$ of the web pages in $\Val$, according to which the search results are displayed to Alice. This preference ordering is based on the ranking of the web content according to relevance and quality of the provided information \cite{page98pagerank,PavlovicD:CSR08}.

\subsection{Composing privacy protocols} 
Security protocols are often composed \cite{PavlovicD:dist06,PavlovicD:ICDCIT12,PavlovicD:CWSP09}. The goal of security protocol composition is to preserve and conjoin the security properties of the protocol components \cite{PavlovicD:CSFW05,PavlovicD:MFPS03,PavlovicD:CSFW03}. The problem is that security properties are usually not compositional  \cite{PavlovicD:JCS05,PavlovicD:JCS04}. The problem with privacy protocols is even more subtle, and in fact intentionally mind-boggling. The goal of privacy protocol composition is often to combine the low-level privacy protocol components in such a way that their privacy properties are subverted, and sublimated into different privacy properties. Alice's privacy goals of the low-level protocol components are replaced by Eve's privacy goals of high-level composite protocols. We spell out some examples, and make a first couple of steps towards analyzing this protocol knot.

\paragraph{Example 3 continued:}
In general, Bob does not own the search engine, but only his own web pages: his resource $\reso\bob:\Var\tores\Val$ is just a map from $\Var_\bob\subseteq \Var$ (which can be thought of as the content menu of his web site, filtered by $\pi_\bob:\Var\pfn\Var$ from the universe of keywords $\Var$) to $\Val$, containing the actual content of the pages that he runs. When Alice visits Bob's web page and submits request $\req^{\alice\bob}$ for some content from his site, he responds according to his policy $\poli^{\alice\bob}$, and displays $\reso{\alice\bob} = \poli^{\alice\bob}\reso\bob\req^{\alice\bob}$. Moreover, in addition to providing the content that Alice explicitly requested, Bob can also try to sell a fragment of Alice's expressed interest to  advertisers, and provide her with some content that she did not explicitly request, but may be related. The search for the advertisers willing to buy a fragment of Alice's interest is a second use that the owner of the search engine, whom we call Gogol\footnote{Nikolai Vasilievich Gogol was a XIX century Russian writer. Gogols are also the ape-like enemies in the video game \emph{Xenoblade Chronicles}.}, will find for his web index $\gamma:\Var\tores\Val$. He will thus extract from $\gamma$ two different resources: 
\begin{itemize}
\item the search index $\reso{\gogol_{se}}:\Var\tores\Val$, assigning to each search term from $\Var$ a source element of web contents from $\Val$, but this time not informative but advertising contents; and on the other hand
\item the advertising index $\reso{\gogol_{ad}}:\ZZZ\tores\Var$, assigning to each advertising opportunity\footnote{Gogol receives advertising requests in a separate privacy protocol. It will be briefly discussed in the next section.} a source element of related search terms.
\end{itemize} 
These two resources are used as follows.
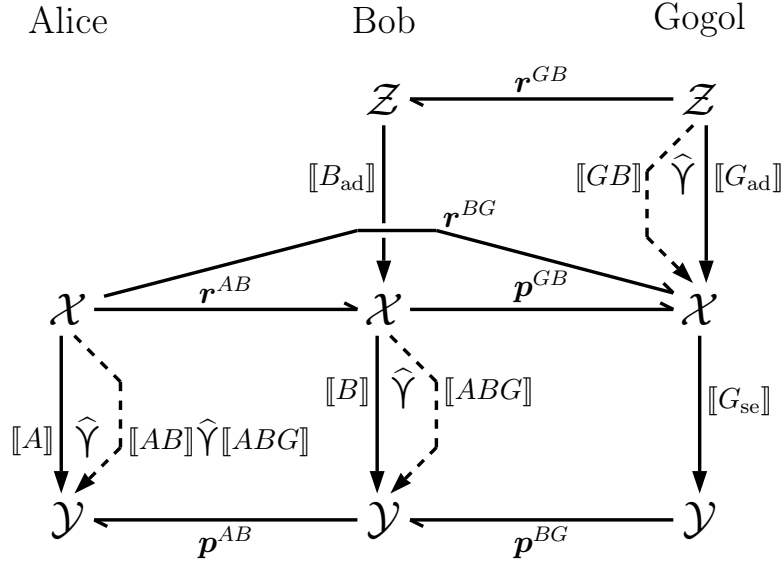
\begin{figure}[h!t]
\centering
\newcommand{\Alce}{\large Alice}
\newcommand{\Exex}{\large Gogol}
\newcommand{\Vars}{{\Large$\Var$}}
\newcommand{\Vals}{{\Large$\Val$}}
\newcommand{\Orig}{{\Large$\ZZZ$}}
\newcommand{\Bobo}{\large Bob}
\newcommand{\Alceresource}{\mbox{\small$\reso{\alice}$}}
\newcommand{\Exresource}{\mbox{\small$\reso{\gogol_{\rm se}}$}}
\newcommand{\Exmoney}{\mbox{\small$\reso{\gogol_{\rm ad}}$}}
\newcommand{\ExBobmoney}{\mbox{\small$\reso{\gogol\bob}$}}
\newcommand{\Alcebobsresource}{\mbox{\small$\reso{\alice\bob}\Fusion\reso{\alice\bob\gogol}$}}
\newcommand{\ABobExperianresource}{\mbox{\small$\reso{\alice\bob\gogol}$}}
\newcommand{\Bobsresource}{\mbox{\small$\reso{\bob}$}}
\newcommand{\Bobcontract}{\mbox{\small$\reso{\bob_{\rm ad}}$}}
\newcommand{\Request}{\req^{\alice\bob}}
\newcommand{\Response}{\poli^{\alice\bob}}
\newcommand{\RequestBobEx}{$\poli^{\gogol\bob}$}
\newcommand{\ResponseBobEx}{$\poli^{\bob\gogol}$}
\newcommand{\RequestExBob}{$\req^{\gogol\bob}$}
\newcommand{\ResponseExBob}{$\req^{\bob\gogol}$}
\newcommand{\addjoin}{\Fusion}
\def\JPicScale{.7}
\ifx\JPicScale\undefined\def\JPicScale{1}\fi
\psset{unit=\JPicScale mm}
\psset{linewidth=0.3,dotsep=1,hatchwidth=0.3,hatchsep=1.5,shadowsize=1,dimen=middle}
\psset{dotsize=0.7 2.5,dotscale=1 1,fillcolor=black}
\psset{arrowsize=1 2,arrowlength=1,arrowinset=0.25,tbarsize=0.7 5,bracketlength=0.15,rbracketlength=0.15}
\begin{pspicture}(0,0)(137.5,100)
\psline[linewidth=0.65](20,45)(70,45)
\psline[linewidth=0.65](20,5)(70,5)
\psline[linewidth=0.65,arrowsize=1.55 2,arrowlength=1.5,arrowinset=0]{->}(73.75,40)(73.75,10)
\rput(15,100){\Alce}
\rput(75,100){\Bobo}
\rput[b](45,46.25){$\Request$}
\rput[t](45,3.75){$\Response$}
\rput[r](72.5,29.38){$\Bobsresource$}
\psline[linewidth=0.65,linestyle=dashed,dash=2 2](25,31.25)(25,18.75)
\rput[r](12.5,20){$\Alceresource$}
\psline[linewidth=0.65,linestyle=dashed,dash=2 2](16.25,40)(25,31.25)
\psline[linewidth=0.65,linestyle=dashed,dash=2 2,arrowsize=1.5 2,arrowlength=1.5,arrowinset=0.05]{->}(25,18.75)(16.25,10)
\psline[linewidth=0.65,arrowsize=1.55 2,arrowlength=1.5,arrowinset=0]{->}(13.75,40)(13.75,10)
\psline[linewidth=0.65](67.5,46.25)(70,45)
\psline[linewidth=0.65](20,5)(22.5,3.75)
\rput[l](26.25,20.62){$\Alcebobsresource$}
\rput(75,45){\Vars}
\rput(15,5){\Vals}
\rput(75,5){\Vals}
\rput(15,45){\Vars}
\rput(18.75,20.62){$\addjoin$}
\psline[linewidth=0.65](80,45)(130,45)
\psline[linewidth=0.65](80,5)(130,5)
\psline[linewidth=0.65,arrowsize=1.55 2,arrowlength=1.5,arrowinset=0]{->}(135,40)(135,10)
\psline[linewidth=0.65,linestyle=dashed,dash=2 2](85,31.25)(85,18.75)
\psline[linewidth=0.65,linestyle=dashed,dash=2 2](76.25,40)(85,31.25)
\psline[linewidth=0.65,linestyle=dashed,dash=2 2,arrowsize=1.5 2,arrowlength=1.5,arrowinset=0.05]{->}(85,18.75)(76.25,10)
\psline[linewidth=0.65](127.5,46.25)(130,45)
\psline[linewidth=0.65](80,5)(82.5,3.75)
\rput(135,45){\Vars}
\rput(135,5){\Vals}
\rput(78.75,30){$\addjoin$}
\rput(135,100){\Exex}
\psline[linewidth=0.65,arrowsize=1.55 2,arrowlength=1.5,arrowinset=0]{->}(75,80)(75,50)
\psline[linewidth=0.65,arrowsize=1.55 2,arrowlength=1.5,arrowinset=0]{->}(136.25,80)(136.25,50)
\psline[linewidth=0.65](80,85)(130,85)
\psline[linewidth=0.65](80,85)(82.5,83.75)
\rput(135,85){\Orig}
\rput(75,85){\Orig}
\rput[b](105,46.25){\RequestBobEx}
\rput[t](105,3.75){\ResponseBobEx}
\rput[bl](86.25,60.62){\ResponseExBob}
\rput[b](105,86.25){\RequestExBob}
\rput[l](136.25,27.5){$\Exresource$}
\rput[l](137.5,70){$\Exmoney$}
\psline[linewidth=0.65,linestyle=dashed,dash=2 2](125,71.25)(125,58.75)
\psline[linewidth=0.65,linestyle=dashed,dash=2 2](133.75,80)(125,71.25)
\psline[linewidth=0.65,linestyle=dashed,dash=2 2,arrowsize=1.5 2,arrowlength=1.5,arrowinset=0.05]{->}(125,58.75)(133.75,50)
\rput(131.88,70.62){$\addjoin$}
\rput[r](73.75,70){$\Bobcontract$}
\rput[r](124.38,70){$\ExBobmoney$}
\rput[l](86.25,29.38){$\ABobExperianresource$}
\psline[linewidth=0.65](21.88,47.5)(70,60)
\psline[linewidth=0.65,border=1.05](70,60)(85,60)
\psline[linewidth=0.65](85,60)(130,48.12)
\psline[linewidth=0.65](128.12,50)(130,48.12)
\end{pspicture}
\caption{Composite privacy protocol: Targeted Advertising (TAd)}
\label{fig:ad}
\end{figure}
Gogol initiates pooling of his and Bob's resources by offering in $\req^{\gogol\bob}$ some advertising opportunities from $\ZZZ$. Bob creates some advertising space $\reso{\bob_{ad}}:\ZZZ\tores \Var$, ready to be inserted into his web site $\reso{\bob}:\Var\tores\Val$, and accepts Gogol's collaboration proposal by $\poli^{\gogol\bob}$. Gogol processes the provided part $\reso{\gogol\bob} = \poli^{\gogol\bob}\reso{\bob_{ad}}\req^{\gogol\bob}$ of Bob's resource, and uses $\reso{\gogol\bob} \Fusion \reso{\gogol_{ad}}$ to determine which of the advertising opportunities from $\ZZZ$ will best suit Bob's site. When Bob receives Alice's request $\req^{\alice\bob}$, he forwards it to Gogol as his request $\req^{\gogol\bob} = \req^{\bob\gogol}\req^{\alice\bob}$ for the ad content. Using the current search index $\reso{\gogol_{se}}$ of  advertising contents and the index $\reso{\gogol\bob} \Fusion \reso{\gogol_{ad}}$ of advertising topics suitable for Bob's site, Gogol provides Bob with the targeted web ad $\reso{\alice\bob\gogol}=\poli^{\alice\bob}\poli^{\bob\gogol}\reso\gogol \req^{\bob\gogol}\req^{\alice\bob}$, deemed to be of interest for Alice because of her interest in Bob's web content. Bob then displays $\reso\bob \Fusion \reso{\alice\bob\gogol}$, and Alice receives $\reso{\alice\bob}\Fusion\reso{\alice\bob\gogol}$. This composite protocol is displayed in the bottom row of Fig.~\ref{fig:ad}. The protocol component where Gogol pays Bob for displaying the ad is omitted, as is the component where the advertiser pays Gogol, and the one where Alice requests to purchase the advertised goods, and the one where she remits the payment to the merchant, etc. We shall see some such protocols in the next section, but the mosaic of protocols for sharing and trading private resources always spreads beyond the horizon.

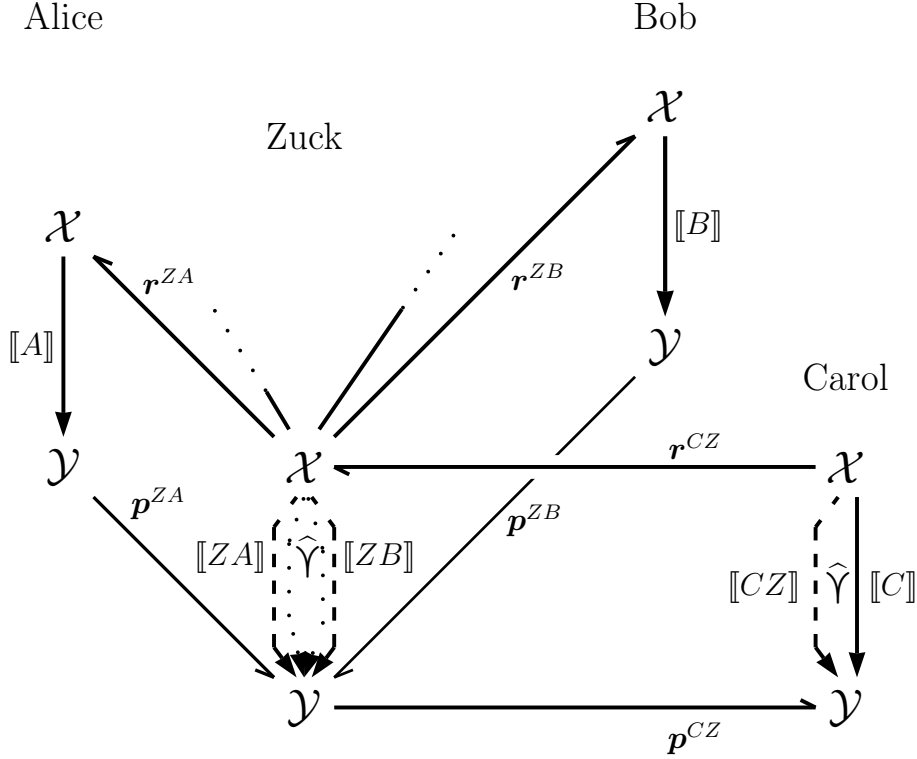
\begin{figure}[h!t]
\centering
\newcommand{\Vars}{{\Large$\Var$}}
\newcommand{\Vals}{{\Large$\Val$}}
\newcommand{\Alce}{\large Alice}
\newcommand{\Bobo}{\large Bob}
\newcommand{\Carol}{\large Carol}
\newcommand{\Zucker}{\large Zuck}
\newcommand{\AlceresourceR}{\reso{\alice\bob}}
\newcommand{\Bobsresource}{\resoo{\bob}1}
\newcommand{\Alceresource}{\reso{\alice}}
\newcommand{\Carolresource}{\reso{\carol\zuck}}
\newcommand{\DaveresourceR}{\reso{\bob}}
\newcommand{\CarolresourceR}{\reso{\carol}}
\newcommand{\BobresourceR}{\resoo{\bob}1}
\newcommand{\ZuckresourceAlice}{\reso{\zuck\alice}}
\newcommand{\ZuckresourceBob}{\reso{\zuck\bob}}
\newcommand{\RequestZuckAlice}{\req^{\zuck\alice}}
\newcommand{\RequestZuckBob}{\req^{\zuck\bob}}
\newcommand{\PoliZuckAlice}{\poli^{\zuck\alice}}
\newcommand{\PoliZuckBob}{\poli^{\zuck\bob}}
\newcommand{\ResponseCD}{\req^{\carol\zuck}}
\newcommand{\Response}{\poli^{\carol\zuck}}
\newcommand{\union}{\Fusion}
\def\JPicScale{.8}
\ifx\JPicScale\undefined\def\JPicScale{1}\fi
\psset{unit=\JPicScale mm}
\psset{linewidth=0.3,dotsep=1,hatchwidth=0.3,hatchsep=1.5,shadowsize=1,dimen=middle}
\psset{dotsize=0.7 2.5,dotscale=1 1,fillcolor=black}
\psset{arrowsize=1 2,arrowlength=1,arrowinset=0.25,tbarsize=0.7 5,bracketlength=0.15,rbracketlength=0.15}
\begin{pspicture}(0,0)(133.75,120)
\psline[linewidth=0.65,arrowsize=1.5 2,arrowlength=1.5,arrowinset=0.05]{->}(131.88,40)(131.88,10)
\rput(0,120){\Alce}
\rput(100,120){\Bobo}
\psline[linewidth=0.65,linestyle=dashed,dash=2 2](125,35)(125,15)
\rput[r](-1.25,65){$\Alceresource$}
\psline[linewidth=0.65,linestyle=dashed,dash=1.2 2.8](128.75,40)(125,35)
\psline[linewidth=0.65,linestyle=dashed,dash=1.2 2.8,arrowsize=1.5 2,arrowlength=1.5,arrowinset=0.05]{->}(125,15)(128.75,10)
\psline[linewidth=0.65,border=1.5](5,80)(35,50)
\psline[linewidth=0.55](5,40)(35,10)
\rput[r](122.5,25){$\Carolresource$}
\rput(130,60){\Carol}
\rput[l](133.75,25){$\CarolresourceR$}
\psline[linewidth=0.65,arrowsize=1.5 2,arrowlength=1.5,arrowinset=0.05]{->}(0,80)(0,50)
\rput(40.62,30){$\union$}
\psline[linewidth=0.7,arrowsize=1.5 2,arrowlength=1.5,arrowinset=0.05]{->}(100,100)(100,70)
\rput[l](101.25,85){$\DaveresourceR$}
\psline[linewidth=0.45](45,10)(95,60)
\psline[linewidth=0.65,border=1.5](45,50)(95,100)
\psline[linewidth=0.65,linestyle=dashed,dash=2 2](35,35)(35,15)
\psline[linewidth=0.65,linestyle=dashed,dash=1.2 2.8](38.75,40)(35,35)
\psline[linewidth=0.65,linestyle=dashed,dash=1.2 2.8,arrowsize=1.5 2,arrowlength=1.5,arrowinset=0.05]{->}(35,15)(38.75,10)
\psline[linewidth=0.65,linestyle=dashed,dash=2 2](45,35)(45,15)
\psline[linewidth=0.65,linestyle=dashed,dash=1.2 2.8](41.25,40)(45,35)
\psline[linewidth=0.65,linestyle=dashed,dash=1.2 2.8,arrowsize=1.5 2,arrowlength=1.5,arrowinset=0.05]{->}(45,15)(41.25,10)
\rput(0,45){\Vals}
\rput(0,85){\Vars}
\rput(40,5){\Vals}
\rput(40,45){\Vars}
\rput(100,65){\Vals}
\rput(100,105){\Vars}
\rput(130,5){\Vals}
\rput(130,45){\Vars}
\psline[linewidth=0.65,border=1.65](45,45)(125,45)
\rput[t](105,2.5){$\Response$}
\psline[linewidth=0.65](45,45)(47.5,43.75)
\psline[linewidth=0.45](45,10)(48.12,11.25)
\psline[linewidth=0.55](31.88,11.25)(35,10)
\psline[linewidth=0.65](91.88,98.75)(95,100)
\psline[linewidth=0.65](5,80)(8.12,78.75)
\psline[linewidth=0.65](45,5)(125,5)
\rput(128.75,25.62){$\union$}
\psline[linewidth=0.65](122.5,6.25)(125,5)
\psline[linewidth=0.65](56.25,71.25)(42.5,51.25)
\psline[linewidth=0.65](33.75,57.5)(37.5,51.25)
\psline[linewidth=0.65,linestyle=dotted,dotsep=3](65,83.75)(56.25,71.25)
\psline[linewidth=0.65,linestyle=dotted,dotsep=3](25,71.25)(33.75,57.5)
\psline[linewidth=0.65,linestyle=dotted,dotsep=3](37.5,32.5)(37.5,20)
\psline[linewidth=0.65,linestyle=dotted,dotsep=3](40,40)(37.5,32.5)
\psline[linewidth=0.65,linestyle=dotted,dotsep=3,arrowsize=1.5 2,arrowlength=1.5,arrowinset=0.05]{->}(37.5,20)(40,10)
\psline[linewidth=0.65,linestyle=dotted,dotsep=3,arrowsize=1.5 2,arrowlength=1.5,arrowinset=0.05]{->}(43.12,18.75)(40,10)
\psline[linewidth=0.65,linestyle=dotted,dotsep=3](43.12,31.25)(43.12,18.75)
\psline[linewidth=0.65,linestyle=dotted,dotsep=3](40.62,40)(43.12,31.25)
\rput(40,100){\Zucker}
\rput[b](105,46.25){$\ResponseCD$}
\rput[l](46.25,30){$\ZuckresourceBob$}
\rput[r](33.75,30){$\ZuckresourceAlice$}
\rput[tl](74.38,78.75){$\RequestZuckBob$}
\rput[bl](13.12,73.75){$\RequestZuckAlice$}
\rput[bl](11.25,36.25){$\PoliZuckAlice$}
\rput[tl](73.75,38.75){$\PoliZuckBob$}
\end{pspicture}
\caption{Composite privacy protocol: Social Networking (SNet)}
\label{fig:social}
\end{figure}

\paragraph{Example 4:} While Gogol the search engine builds ranked indices of the content existing on the web, and shares these resources with Alice, Bob, and Carol, Zuck the social engine elicits the content from Alice, Bob, and Carol, builds an index of that, and shares their content with them as their social interactions. A bird's eye view of a fragment of this process is displayed in Fig.~\ref{fig:social}.
Zuck's requests $\req^{\zuck\subj}$ to all $\subj\in \Nod$ offer to index and distribute everyone's media and contents. In response he receives $\reso{\zuck\subj} = \poli^{\zuck\subj}\reso\subj\req^{\zuck\subj}$ from all $\subj\in \Nod$, which are some of their private media and content that they want to share with their social network. To better distribute them, Zuck conjoins the shared resources into his main resource, the social engine index 
\[ 
\zeta\   \ =\ \ \reso{\zuck_{se}} \ \ = \ \ \Fusion_{\subj\in \Nod} \reso{\zuck\subj}\ \ =\ \ 
\Fusion_{\subj\in \Nod} \poli^{\zuck\subj}\reso\subj\req^{\zuck\subj}\]
and shares a part of it with Alice 
\[ \reso{\alice\zuck}\ \  =\ \  \poli^{\alice\zuck}\reso{\zuck_{se}} \req^{\alice\zuck}
%\ \ =\ \  \poli^{\alice\zuck}\left(\Fusion_{\subj\in \Nod} \reso{\zuck\subj}\right) \req^{\alice\zuck}
\ \ =\ \ \poli^{\alice\zuck}\left(\Fusion_{\subj\in \Nod} \poli^{\zuck\subj}\reso\subj\req^{\zuck\subj}\right) \req^{\alice\zuck}
\]
The idea is that Zuck's resource $\reso{\alice\zuck}$  shared with Alice consists of Alice's friends' private contents processed by Zuck. As a participant who only relays messages, Zuck plays the role of a legitimate Man-in-the-Middle in this privacy protocol. The Man-in-the-Middle pattern is usually an attack strategy, and not a protocol role. Protocols sometimes use a Trusted Third Party for a particular functionality.  However, Zuck is here more than a Trusted Third Party (TTP), because he does not provide a particular functionality, but processes \emph{all}\/ protocol content; he is also less than a TTP, because he does not originate any content. Most importantly, he fuses all protocol content into his private resource $\zeta= \reso{\zuck_{se}}:\Var\tores \Val$. Like in a search  engine, the weights of the source elements $\zeta_x \in \Dis\Val$ determine  a preference ranking of the content. 
%which users will receive which of the content accessible to them they will see, and in which order. More precisely, 

Zuck's privacy policy for Carol $\poli^{\carol\zuck}$ determines which of Alice's, Bob's and Dave's postings Carol will see, and in which order. Carol's request $\req^{\carol\zuck}$ can filter out what she does not want to see; but Carol cannot use $\req^{\carol\zuck}$ to request what she will see, because Zuck's index is not available to her. %she passively receives what Zuck provides. 

Zuck's resource $\reso{\zuck_{se}}$ and all of his policies $\poli^{\subj\zuck}$ also take into account Alice's and Bob's privacy policies $\poli^{\zuck\alice}$ and $\poli^{\zuck\bob}$, which may specify whether Carol should have access to their content. However, the existing social engines provide policy languages for $\poli^{\zuck\subj}$ at the expressiveness level of 1960s operating systems, before the concepts of access control were introduced, where only a small number of fixed access control policies are available. The users are usually offered to either make their postings available to general public, or to all of their private contacts ("friends"), and sometimes also to all of their contacts' contacts ("friends' friends"). The option of establishing subgroups and hierarchies of private contacts is not offered. The option of sharing content with specific subgroups is offered as a separate functionality, but the recipients must be reentered with each message, or kept in an address book, which is separate from privacy policies. 

While more expressive access control systems, allowing refined privacy policies, would surely vastly enhance usability of the social engine for Alice and Bob, allowing them to manage their content using the access control mechanisms that they got used to while using their computers, this would leave Zuck with narrower choices of his policies. And Zuck, of course, monetizes his services by manipulating the weight biases in his social engine index $\reso{\zuck_{se}} : \Var \tores \Val$, and by selling the choices what to display to each user. These choices are expressed by his policy $\poli^{\subj\zuck}:\Val\tores\Val$ for each user $\subj\in \Nod$. The more privacy choices are made by Alice and Bob as content originators, the fewer privacy choices are left to Zuck as the content aggregator and monetizer.
 
%It provides users' content according to users' privacy policies. But since there is a lot of content, it can choose in which order, or with which preferences to serve these source elements to the users: the order in which users see their friends' posts are determined by the social engine. The engine Zuck then sells this service, the Policy part of the RP protocol. If you get lots of likes, you influence your friends.
%

\section{Local privacy failures: Covert channels}
\label{Sec:nonint}
% !TEX root = 0-PrivT.tex
\subsection{Privacy protocol failures}
A privacy protocol establishes an \emph{overt}\/ private channel to distribute private resources. In the basic RP-protocol in Fig.~\ref{Fig:privproto}, Bob releases some of his private resource to Alice. In the composite privacy protocols, multiple resources can be shared or exchanged between multiple subjects along multiple overt channels. Privacy protocol failures occur when, in addition to the overt channels, some \emph{covert}\/ channels are also established. In the simplest case, a covert channel from Bob to Alice arises when Bob intends to release to Alice the part $\reso{\alice\bob} = \poli^{\alice\bob}\reso{\bob} \req^{\alice\bob} $ of his resource $\reso\bob$, but Alice manages to extract in $\reso\alice\fusion\reso{\alice\bob}$ a bigger part of $\reso\bob$. Bob's privacy problem with respect to Alice is thus to constrain his policy $\poli^{\alice\bob}$ to assure that $\reso{\alice\bob} = \poli^{\alice\bob}\reso{\bob} \req^{\alice\bob} $ satisfies the requirement
\bea\label{eq:bob-alice-nonint}
\Big(\reso\alice \fusion\reso{\alice\bob} \Big)\dmeet \reso\bob & \prec & \reso{\alice\bob} 
\eea
Note that \eqref{eq:bob-alice-nonint} does not require that Alice is prevented from extracting some of Carol's resource from the interaction with Bob, which she could achieve by cross-referencing $\reso{\alice\bob}$ with some information about $\reso\carol$ previously stored in $\reso\alice$. Requirement \eqref{eq:bob-alice-nonint} also does not require that Dave is prevented from extracting some of Bob's resource from a later interaction with Alice, who may make some of $\reso{\alice\bob}$ available to him. Requirement \eqref{eq:bob-alice-nonint} just expresses Bob's \emph{local}\/ privacy goal, concerning the present channel between him and Alice. The stronger requirement 
\bea\label{eq:bob-alice-strong}
\reso\alice \fusion\reso{\alice\bob} & \prec & \reso{\alice\bob} 
\eea
would, of course, be more socially responsible; but Bob only has access to $\reso\bob$, no direct access to $\reso\alice$, and he has therefore no way to observe or preclude any crossreferences that Alice might be able establish between $\reso\alice$ and $\reso{\alice\bob}$, except those that he observes in $\reso\bob$. If Bob could enforce \eqref{eq:bob-alice-strong}, then he could realizing the Dalenius' desideratum, discussed in Sec.~\ref{Sec:Intro}. The non-local privacy protocol failures, resulting from \emph{non-local covert channels}, will be discussed in the next section. In this section, our considerations are limited to \emph{local covert channels}\/ that may arise from Bob to Alice, in parallel with the overt channel set up by an RP-protocol like in Fig.~\ref{Fig:privproto}.

But even for this, there is more to the story than \eqref{eq:bob-alice-nonint}. A covert channel may arise even if there is no overt channel at all. Take the trivial case, where Bob's policy is to not share anything with Alice, which is expressed by the everywhere undefined function $\poli^{\alice\bob}=\emptyset$ giving $\reso{\alice\bob} = \emptyset$. Requirement \eqref{eq:bob-alice-nonint} thus boils down to
\bear
\reso\alice \dmeet \reso\bob & \prec & \emptyset 
\eear
This is easily assured by $\Val_\alice \cap \Val_\bob = \emptyset$ and $\Var_\alice \cap \Var_\bob = \emptyset$. But if a resource is shared, then the users may submit their inputs in turns, get their outputs in turns, yet there will be \emph{interferences}.

\paragraph{Example of interference.} Suppose that Alice, Bob and Carol live in the same highrise building and share an elevator. Each of them inputs an elevator call, and then each of them receives the elevator service. During the service, the elevator is not available. After each service, the elevator cabin may be left at a different floor. By observing the state of the elevator (its location, whether it is currently moving, etc.), Alice can obtain information about Bob's and Carol's interactions with the elevator; by interfering with its state, she can make it unavailable for them. The elevator can thus provide a covert channel between Alice, 
Bob and Carol just by the virtue of being shared, without providing an overt channel between them.

\paragraph{Covert channels.} Even if Alice's, Bob's and Carol's private inputs come from disjoint sets $\Var_\alice$, $\Var_\bob$ and $\Var_\carol$ (which is the case if the elevator authenticates calls, say by requiring room cards in a hotel, or biometric checks in an appartment building), the state of the elevator in general depends on a whole history of private inputs. Suppose that each user $\subj\in \Nod$ has a devoted set of actions $\Sigma_\subj$, so that only Alice can call the elevator to her penthouse, because only $\Sigma_\alice$ contains that action. If $\Sigma_\alice \cap \Sigma_\bob = \emptyset$, then it follows that there is no overt channel between Alice and Bob, because all sequences of inputs that Alice alone or Bob alone may enter are also disjoint, i.e. $\Var_\alice \cap\Var_\bob = \emptyset$  also holds for
\bear
\Var_\subj & = & \Sigma_\subj^\ast
\eear
But the actual inputs that the elevator receives are usually not from any user alone, but they are shuffled in
\bear
\Var & = & \left(\coprod_{\subj \in \Nod} \Sigma_\subj\right)^\ast
\eear
The projections are thus in the form
\[%\beq\label{eq:privproj}
\begin{tikzar}[column sep=3em]
\Var_\alice = \Sigma_\alice^\ast
 \arrow[bend right = 12,tail]{rr}[swap]{\underline\pi_\alice}
 \&\& 
 \left(\coprod_{\subj\in \Nod} \Sigma_\subj\right)^\ast = \Var  
 \arrow[bend right = 12,two heads,dashed%,rightharpoondown
]{ll}[swap]{\overline\pi_\alice} 
\end{tikzar}
\]%\eeq
where 
\begin{itemize}
\item $\overline \pi_\alice(x) \ =\ \begin{cases}
<> & \mbox{ if } x=<>\\
\sigma:: \overline\pi_\alice (x')  & \mbox{ if } x = \sigma::x' \mbox{ and } \sigma \in \Sigma_\alice\\
\overline\pi_\alice (x')  & \mbox{ if } x = \sigma::x' \mbox{ and } \sigma \not \in \Sigma_\alice
\end{cases}$
\item $\underline\pi_\alice = \iota^\ast_\alice$ lifts the inclusion $\iota_\alice : \Sigma_\alice\inclusion \coprod_{\subj\in \Nod} \Sigma_\subj$  to the sequences.
\end{itemize}
An interference of Alice's use of the elevator with Bob's and Carol's use occurs if Alice observes two situations  $x,x' \in \Var$ such that
\begin{itemize}
\item $\pi_\alice(x) = \pi_\alice(x')$ --- her own inputs are the same, but
\item $\rho_\alice\reso \alice ( x)\neq \rho_\alice\reso \alice (x')$ --- the elevator produces different outputs, 
\end{itemize}
where $\rho_\alice = \underline\rho_\alice \overline\rho_\alice :\Val \pfn \Val$ is the projector from \eqref{eq:privproj}. The difference must be caused by Bob's or Carol's different inputs within the strings $x$ and $x'$. Their inputs are not directly observable for Alice, and she may not know what they are, or who has entered them; but the fact that they are different gives Alice a single bit of information about Bob's and Carol's actions. Hence the covert channel.

\paragraph{Noninterference.} The task of recognizing such covert channels and the methods to eliminate them have been extensively studied in system security, and led to the requirement of \emph{noninterference}, and a whole range of related properties \cite{goguen-meseguer,McCullough:CSFW88,McLean,McLean:encyclopedia,RushbyJ:nonint,Sutherland:nondeducibility}. A simple way to state the noninterference requirement is in terms of  the equivalence relations 
\bear
x \ineq{\alice} x' & \iff & 
\pi_\alice(x) = \pi_\alice(x') \mbox{ and}\\
x \oueq{\alice} x' & \iff & 
\rho_\alice \reso{\alice} (x) = \rho_\alice\reso\alice (x') 
\eear
The noninterference requirement is then simply
\bea\label{eq:noninter}
x \ineq{\alice} x' & \implies &
x \oueq{\alice} x' 
\eea
This is a very strong requirement: that $\reso\alice$ does not interfere with anyone else's inputs or outputs, except Alice's. 
The version localized to Alice and Bob is obtained by defining 
\bear
x \ineq{\alice\bob} x' & \iff & 
\rho_\alice \reso{\alice\bob} (x) = \rho_\alice\reso{\alice\bob} (x')\\
x \oueq{\alice\bob} x' & \iff & 
\rho_\alice \Big(\left(\reso\alice \fusion\reso{\alice\bob} \right)\dmeet \reso\bob\Big)(x)\  =
%\\
%&& %
\   
\rho_\alice\Big(\left(\reso\alice \fusion\reso{\alice\bob} \right)\dmeet \reso\bob\Big) (x') 
\eear 
and requiring
\bear
x \ineq{\alice\bob} x' & \implies &
x \oueq{\alice\bob} x' 
\eear
For deterministic resources, i.e. when $\reso \alice$, $\reso{\alice\bob}$ etc. are partial functions $\Var\pfn \Val$, it is easy to see that e.g. requirement \eqref{eq:noninter} is equivalent to the existence of a function $\overline{\reso\alice}$ that makes the diagram on the left-hand side of  \eqref{eq:privproj} commute.
%\begin{figure}[H!t]
%\begin{center}
\begin{gather}\label{eq:nonint-diag}
\begin{tikzar}[row sep=4em,column sep=6em]
\Var \ar[two heads]{d}[swap]{\overline\pi_\alice} 
\ar[rightharpoonup]{r}{\reso\alice} \& \Val \ar[two heads]{d}{\overline\rho_\alice} \\
\Var_\alice \ar[rightharpoonup,dashed]{r}
[swap]{\overline{\reso{\alice}}}
\& \Val_\alice
\end{tikzar}\qquad \qquad 
\begin{tikzar}[row sep=4em,column sep=6em]
\Var \ar{d}[swap]{\pi_\alice} \ar[rightharpoonup]{r}{\reso\alice}
 \ar{dr}[description]{{\reso{\alice}}} 
\& \Val \ar{d}{\rho_\alice} \\
\Var \ar[rightharpoonup]{r}[swap]{{\reso{\alice}}} 
\& \Val
\end{tikzar}\end{gather}
Recalling the projections from \eqref{eq:privproj}, it is clear that the commutativity of the diagram on the right-hand side of \eqref{eq:nonint-diag} also  provides an equivalent form of the noninterference requirement. Yet another equivalent form is that $\reso\alice = \rho_\alice\reso\alice\pi_\alice$.

For general resources from Def.~\ref{Def:resource}, the requirements from \eqref{eq:nonint-diag} can still be imposed, but validating that the probability distributions of source elements are equal on the nose is almost never feasible in practice. Requiring that diagrams \eqref{eq:nonint-diag} commute up to $\varepsilon$ takes us in the direction of \emph{differential privacy} \cite{dwork2014}, which is a popular and practical tool for dealing with covert channels. In the context of  privacy protocols, its effectiveness, at least in the original form, seems to be limited to \emph{local}\/ covert channels. --- But what are non-local covert channels? 

\section{Nonlocal privacy failures: Covert protocols}
\label{Sec:vulnerabilities}
% !TEX root = 0-PrivT.tex

The rapid transformation of the socio-technical context of privacy is an acute socio-political problem,
%. Although this problem seems easier to avoid in research than in everyday life, it has been 
which has been described from many angles \cite{acquisti2007digital,angwin2014dragnet,ball2012routledge,diffie2010privacy,zuboff2019age}. 
The idea of this research is to shed some light on the problem of privacy by analyzing privacy protocol problems.
In this section, we
% try to shed light on the problem of privacy on a couple of typical privacy protocol problems. 
attempt to outline the unusual shape of some of the typical privacy protocol problems. They are unusual in that their vulnerabilities are not %do not come from 
 design flaws that open some attack vectors for outside attackers, as they do in security protocols. Typical privacy protocol vulnerabilities seem inherent to protocols themselves. Not that the protocols always have back doors; but they have transparent roofs. They are not vulnerable to subversions, or level-below attacks, but to sublimations, or level-above applications.

\subsection{Man-in-the-Middle protocols}\label{sec:mitm}
We have seen that Zuck was an MitM in the SNet protocol on Fig.~\ref{fig:social}, where his private resource $\reso{\zuck_{se}}$ is inserted in-between everybody else's private resources, as their fusion; and that Bob is an MitM in the TAd-protocol on Fig.~\ref{fig:ad}, where he is inserted in-between Gogol and Alice, and serves Gogol's targeted ads to Alice. A level above SNet and TAd, both Zuck and Gogol are MitMs in the privacy protocols where they monetize their services (social networking and web search, respectively) by inserting themselves in-between the advertisers and their targets (usually voters or consumers). Zuck's protocol for monetizing his social engine through Social Influencing (SInf) is displayed in Fig.~\ref{fig:sublime}. This is a level-above protocol in the sense that it is built on top of the SNet protocol, which it uses as an encapsulated procedure. In Fig.~\ref{fig:sublime}, we only display a single RP-component of SNet: the rectangle on the left, where Zuck provides Carol her friends' postings. We represent SNet by its single component only because we do not have a good diagrammatic notation to encapsulate a low-level protocol as an atom of a high-level protocol.
\begin{figure}[h!t]
\newcommand{\Vars}{{\Large$\Var$}}
\newcommand{\Vals}{{\Large$\Val$}}
\newcommand{\Orig}{{\Large$\ZZZ$}}
\newcommand{\Money}{{\LARGE$\$$}}
\newcommand{\query}{$\req^{\carol\zuck}$}
\newcommand{\retrieve}{$\poli^{\carol\zuck}$}
\newcommand{\indx}{$\reso{\zuck_{se}}$}
\newcommand{\indxPre}{$\reso{\zuck_{ad}}$}
\newcommand{\indxPost}{$\reso{\zuck_{bu}}$}
\newcommand{\IndxMoney}{$\reso{\zuck_{\$}}$}
\newcommand{\indxxMoney}{$\reso{\tizer_{\$}}$}
\newcommand{\indxxx}{$\reso{\zuck_{\$}\tizer_{\$}}$}
\newcommand{\indxx}{$\reso{\tizer_{ad}}$}
\newcommand{\indxxRec}{$\reso{\tizer\zuck}$}
\newcommand{\advert}{$\req^{\tizer\zuck}$}
\newcommand{\supply}{$\reso\carol$}
\newcommand{\monetize}{$\poli^{\tizer\zuck}, \req^{\zuck\tizer}$}
\newcommand{\monetizePoli}{$\poli^{\zuck\tizer}$}
\newcommand{\user}{\large Carol}
\newcommand{\SE}{\large Zuck}
\newcommand{\advertiser}{\large Tizer}
\newcommand{\union}{\Fusion}
\newcommand{\addmoney}{+}
\begin{center}
\def\JPicScale{.8}
%\hspace{-3em}
\ifx\JPicScale\undefined\def\JPicScale{1}\fi
\psset{unit=\JPicScale mm}
\psset{linewidth=0.3,dotsep=1,hatchwidth=0.3,hatchsep=1.5,shadowsize=1,dimen=middle}
\psset{dotsize=0.7 2.5,dotscale=1 1,fillcolor=black}
\psset{arrowsize=1 2,arrowlength=1,arrowinset=0.25,tbarsize=0.7 5,bracketlength=0.15,rbracketlength=0.15}
\begin{pspicture}(0,0)(96.88,135)
\psline[linewidth=0.55](10,90)(40,90)
\psline[linewidth=0.55,arrowsize=1.5 2,arrowlength=1.5,arrowinset=0]{->}(45,85)(45,65)
\rput[l](46.25,75){\indx}
\rput[b](24.38,91.25){\query}
\rput[b](25,60.62){\retrieve}
\psline[linewidth=0.55](90,120)(50,120)
\psline[linewidth=0.55](50,30)(90,30)
\psline[linewidth=0.55,arrowsize=1.5 2,arrowlength=1.5,arrowinset=0]{->}(45,115)(45,95)
\psline[linewidth=0.55](40,60)(10,60)
\rput[l](96.88,75.62){\indxx}
\rput[r](2.5,75){\supply}
\rput[t](70,118.75){\advert}
\rput[b](70,31.25){\monetize}
\rput(5,135){\user}
\rput(45,135){\SE}
\rput(95,135){\advertiser}
\psline[linewidth=0.55,arrowsize=1.5 2,arrowlength=1.5,arrowinset=0]{->}(45,55)(45,35)
\psline[linewidth=0.55,arrowsize=1.5 2,arrowlength=1.5,arrowinset=0]{->}(3.75,85)(3.75,65)
\psline[linewidth=0.55,arrowsize=1.5 2,arrowlength=1.5,arrowinset=0]{->}(96.25,115)(96.25,35)
\psline[linewidth=0.55](52.5,118.75)(50,120)
\psline[linewidth=0.55](40.62,90)(38.12,91.25)
\psline[linewidth=0.55](12.5,58.75)(10,60)
\psline[linewidth=0.55](90,30)(87.5,31.25)
\psline[linewidth=0.65,linestyle=dashed,dash=2 2](12.5,78.75)(12.5,71.25)
\psline[linewidth=0.65,linestyle=dashed,dash=2 2](6.25,85)(12.5,78.75)
\psline[linewidth=0.65,linestyle=dashed,dash=2 2,arrowsize=1.5 2,arrowlength=1.5,arrowinset=0.05]{->}(12.5,71.25)(6.25,65)
\psline[linewidth=0.65,linestyle=dashed,dash=2 2](86.25,106.25)(86.25,42.5)
\psline[linewidth=0.65,linestyle=dashed,dash=2 2](93.75,115)(86.25,106.25)
\psline[linewidth=0.65,linestyle=dashed,dash=2 2,arrowsize=1.5 2,arrowlength=1.5,arrowinset=0.05]{->}(86.25,42.5)(93.75,35)
\rput(7.5,76.25){$\union$}
\rput(91.25,76.25){$\union$}
\rput(5,60){\Vals}
\rput(45,60){\Vals}
\rput(5,90){\Vars}
\rput(45,90){\Vars}
\psline[linewidth=0.55,arrowsize=1.5 2,arrowlength=1.5,arrowinset=0]{->}(95,25)(95,5)
\psline[linewidth=0.55](90,0)(50,0)
\psline[linewidth=0.55,arrowsize=1.5 2,arrowlength=1.5,arrowinset=0]{->}(43.75,25)(43.75,5)
\psline[linewidth=0.65,linestyle=dashed,dash=2 2](46.25,25)(52.5,18.75)
\psline[linewidth=0.65,linestyle=dashed,dash=2 2,arrowsize=1.5 2,arrowlength=1.5,arrowinset=0.05]{->}(52.5,11.25)(46.25,5)
\psline[linewidth=0.55](52.5,-1.25)(50,0)
\psline[linewidth=0.65,linestyle=dashed,dash=2 2](52.5,18.75)(52.5,11.25)
\rput(45,120){\Orig}
\rput(95,120){\Orig}
\rput(45,0){\Money}
\rput(95,0){\Money}
\rput(45,30){\Orig}
\rput(95,30){\Orig}
\rput(48.12,15){$\addmoney$}
\rput[r](85,75){\indxxRec}
\rput[l](46.25,105.62){\indxPre}
\rput[l](46.25,45.62){\indxPost}
\rput[r](43.12,15.62){\IndxMoney}
\rput[l](96.25,15.62){\indxxMoney}
\rput[l](53.75,15){\indxxx}
\rput[b](70,1.25){\monetizePoli}
\end{pspicture}
\caption{Level-above privacy protocol: Social Influencing (SInf)}
\label{fig:sublime}
\end{center}
\end{figure}
SInf uses SNet on the left to leverage its business process on the right. Tizer is an advertising campaign manager, and his resource $\reso{\tizer_{ad}}: \ZZZ\tores\ZZZ$ is a campaign strategy, presented as a Markov chain over the a set of campaign messages $\ZZZ$. Tizer initiates an SInf protocol run with $\req^{\tizer\zuck}$, requesting that Zuck places the campaign messages of interest for Tizer at that moment. Zuck runs the request through his advertising index $\reso{\zuck_{ad}}:\ZZZ\tores \Var$, assigning to each campaign message from $\ZZZ$ a suitable context of recipients and their postings in $\Var$, and his social engine $\reso{\zuck_{se}}:\Var\tores \Val$ then inserts the campaign messages among Carol's friends' postings. Zuck's business engine $\reso{\zuck_{bu}}: \Val\tores \ZZZ$ then generates the service report $\poli^{\tizer\zuck}$ and the invoice $\req^{\zuck\tizer}$ for Tizer, who responds with the payment $\poli^{\zuck\tizer}$ from his advertising budget $\reso{\tizer_{\$}}$. The payment $\reso{\zuck\tizer}$ is added to Zuck's budget $\reso{\zuck_{\$}}$.

\subsection{Resource inflation, privacy deflation}\label{sec:alibi}
Dave would like to make pancakes, but he does not have either milk or eggs, so he needs to borrow from Alice, who is his neighbor. But he is shy by nature, and he already borrowed many things. So to avoid intruding into Alice's privacy even more, Dave asks Carol to ask Alice for milk, and he asks Bob to ask her for eggs. If Dave is also short of flour and oil, he can also ask Elizabeth and Frank to knock on Alice's door and ask for that. Then he collects it all, and makes pancakes. This \emph{cross-sharing}\/ privacy protocol is displayed in Fig.~\ref{fig:crossref}.
\begin{figure}[h!t]
\centering
\newcommand{\Alce}{\large Alice}
\newcommand{\Vars}{{\Large$\Var$}}
\newcommand{\Vals}{{\Large$\Val$}}
\newcommand{\Bobo}{\large Bob}
\newcommand{\Carol}{\large Carol}
\newcommand{\Dave}{\large Dave}
\newcommand{\Alceresource}{\reso{\alice}}
\newcommand{\Alcebobsresource}{\reso{\bob\alice}}
\newcommand{\Bobsresource}{\reso{\bob}}
\newcommand{\CarolsResource}{\reso{\carol}}
\newcommand{\CarolAliceResource}{\reso{\carol\alice}}
\newcommand{\Davesresource}{\reso{\dave}}
\newcommand{\DaveBobCarolResource}{\reso{\dave^{(\bob\carol)}\!\!\alice}}
\newcommand{\Request}{\req^{\bob\alice}}
\newcommand{\Response}{\poli^{\bob\alice}}
\newcommand{\ReqDaveBob}{\req^{\dave\bob}_\alice}
\newcommand{\PoliDaveBob}{\poli^{\dave\bob}_\alice}
\newcommand{\ReqDaveCarol}{\req^{\dave\carol}_\alice}
\newcommand{\PoliDaveCarol}{\poli^{\dave\carol}_\alice}
\newcommand{\addjoin}{\Fusion}
\def\JPicScale{.8}
\ifx\JPicScale\undefined\def\JPicScale{1}\fi
\psset{unit=\JPicScale mm}
\psset{linewidth=0.3,dotsep=1,hatchwidth=0.3,hatchsep=1.5,shadowsize=1,dimen=middle}
\psset{dotsize=0.7 2.5,dotscale=1 1,fillcolor=black}
\psset{arrowsize=1 2,arrowlength=1,arrowinset=0.25,tbarsize=0.7 5,bracketlength=0.15,rbracketlength=0.15}
\begin{pspicture}(0,0)(136.25,80)
\psline[linewidth=0.65](20,45)(50,45)
\psline[linewidth=0.65](20,5)(50,5)
\psline[linewidth=0.65,linestyle=dashed,dash=2 2](25,31.25)(25,18.75)
\psline[linewidth=0.65,linestyle=dashed,dash=2 2](16.25,40)(25,31.25)
\psline[linewidth=0.65,linestyle=dashed,dash=2 2,arrowsize=1.5 2,arrowlength=1.5,arrowinset=0.05]{->}(25,18.75)(16.25,10)
\psline[linewidth=0.65,arrowsize=1.55 2,arrowlength=1.5,arrowinset=0]{->}(13.75,40)(13.75,10)
\psline[linewidth=0.65](47.5,46.25)(50,45)
\psline[linewidth=0.65](20,5)(22.5,3.75)
\rput(55,45){\Vars}
\rput(15,5){\Vals}
\rput(55,5){\Vals}
\rput(15,45){\Vars}
\rput(19.38,29.38){$\addjoin$}
\psline[linewidth=0.6](100,65)(130,65)
\psline[linewidth=0.35](100,25)(130,25)
\psline[linewidth=0.6,arrowsize=1.55 2,arrowlength=1.5,arrowinset=0]{->}(135,60)(135,30)
\rput(135,80){\Alce}
\rput(95,80){\Bobo}
\rput[b](115,66.25){$\Request$}
\rput[b](115,26.25){$\Response$}
\psline[linewidth=0.35,linestyle=dashed,dash=2 2](105,51.25)(105,38.75)
\rput[l](136.25,45){$\Alceresource$}
\psline[linewidth=0.35,linestyle=dashed,dash=2 2](96.25,60)(105,51.25)
\psline[linewidth=0.35,linestyle=dashed,dash=2 2,arrowsize=1.5 2,arrowlength=1.5,arrowinset=0.05]{->}(105,38.75)(96.25,30)
\psline[linewidth=0.6](127.5,66.25)(130,65)
\psline[linewidth=0.35](100,25)(102.5,23.75)
\rput[l](106.25,45){$\Alcebobsresource$}
\rput(135,65){\Vars}
\rput(95,25){\Vals}
\rput(135,25){\Vals}
\rput(95,65){\Vars}
\psline[linewidth=0.6](20,46.25)(90,63.12)
\psline[linewidth=0.6](86.88,63.75)(90,63.12)
\psline[linewidth=0.6,border=1.5](61.25,45)(130,61.88)
\psline[linewidth=0.6](126.88,62.5)(130,61.88)
\psline[linewidth=0.6](60,5.62)(128.75,22.5)
\psline[linewidth=0.6](126.88,23.12)(130,22.5)
\psline[linewidth=0.35](20,7.5)(88.75,24.38)
\psline[linewidth=0.35](86.25,25)(89.38,24.38)
\psline[linewidth=0.4,linestyle=dotted,dotsep=2](56.88,53.12)(80.62,58.12)
\psline[linewidth=0.4,linestyle=dotted,dotsep=2](44.38,48.12)(75.62,51.88)
\psline[linewidth=0.35](27.5,46.88)(57.5,53.12)
\psline[linewidth=0.35](29.38,46.25)(59.38,50)
\psline[linewidth=0.35](105,61.88)(121.88,63.12)
\psline[linewidth=0.35](106.88,58.75)(121.88,61.88)
\psline[linewidth=0.4,linestyle=dotted,dotsep=2](96.88,61.25)(106.88,61.88)
\psline[linewidth=0.4,linestyle=dotted,dotsep=2](82.5,53.75)(112.5,60)
\psline[linewidth=0.35](119.38,63.75)(121.88,63.12)
\psline[linewidth=0.35](119.38,61.88)(121.88,61.88)
\psline[linewidth=0.25,linestyle=dotted,dotsep=2](56.25,13.12)(80,18.12)
\psline[linewidth=0.25,linestyle=dotted,dotsep=2](43.75,8.12)(75,11.88)
\psline[linewidth=0.25](26.88,6.88)(56.88,13.12)
\psline[linewidth=0.25](28.75,6.25)(58.75,10)
\psline[linewidth=0.25](104.38,21.88)(121.25,23.12)
\psline[linewidth=0.25](106.25,18.75)(121.25,21.88)
\psline[linewidth=0.25,linestyle=dotted,dotsep=2](96.25,21.25)(106.25,21.88)
\psline[linewidth=0.25,linestyle=dotted,dotsep=2](81.88,13.75)(111.88,20)
\psline[linewidth=0.25](118.75,23.75)(121.25,23.12)
\psline[linewidth=0.25](118.75,21.88)(121.25,21.88)
\rput(55,80){\Carol}
\rput(15,80){\Dave}
\psline[linewidth=0.65,linestyle=dashed,dash=2 2,border=1](65,31.25)(65,18.75)
\psline[linewidth=0.65,linestyle=dashed,dash=2 2,border=1](56.25,40)(65,31.25)
\psline[linewidth=0.65,linestyle=dashed,dash=2 2,border=1,arrowsize=1.5 2,arrowlength=1.5,arrowinset=0.05]{->}(65,18.75)(56.25,10)
\rput[r](12.5,28.12){$\Davesresource$}
\rput[l](26.88,28.12){$\DaveBobCarolResource$}
\rput[l](66.25,25){$\CarolAliceResource$}
\rput[br](45.62,53.12){$\ReqDaveBob$}
\rput[br](45.62,13.75){$\PoliDaveBob$}
\rput[t](35,43.75){$\ReqDaveCarol$}
\rput[t](35,3.75){$\PoliDaveCarol$}
\rput[t](85.62,49.38){$\ReqDaveCarol$}
\rput[t](85.62,10.62){$\PoliDaveCarol$}
\end{pspicture}
\caption{Resource inflation:  Cross-referencing and cross-sharing}
\label{fig:crossref}
\end{figure}
Dave developed this protocol when he worked as a police detective. Alice was often a suspect in his investigations, so whenever he needed to check the details of Alice's alibi, it turned out to be better to ask Alice's friends, than to ask her directly. By conjoining the information obtained from others, he would not only get more details, without drawing attention with too many requests at once, but he could also \emph{cross-reference}\/ the different details that Alice provided to different people. If she tells one person one thing, and another person another thing, then Dave would detect an inconsistency in Alice's alibi. While a fusion of inconsistent sources eliminates the inconsistent parts, and thus contains less information than either of the original sources, the inconsistency itself becomes a \emph{higher-order information}, as mentioned at  the end of Sec.~\ref{Sec:resources}. Level-above attacks and protocols sublimate entire low-level protocol sessions as higher-order information, and manipulate it towards some privacy goals. The loss of a private  resource for Alice is a gain of a private resource for Dave. Having detected an inconsistency in Alice's alibi by asking around in  a cross-referencing protocol, like the alibi  check interpretation of Fig.~\ref{fig:crossref}, Dave confronts Alice himself in a level-above protocol, and requests more higher-order information, to eliminate the  inconsistencies from the acquired resources. An analogous level-above sublimation of the cross-sharing protocol in the pancake sources interpretation of Fig.~\ref{fig:crossref} would be that Dave asks Bob, Carol, Elizabeth and Frank to leave the milk, egg, flour and oil that they got for him from Alice --- to leave if all for him with Alice. Then he could collect it all simply by asking Alice if the neighbors by any chance left anything for him.

The problem of cross-referencing is a well-studied privacy problem, because it arises in statistical databases \cite{SweeneyL:reidentification,Shmatikov:netflix,SweeneyL:weaving}. Statistical databases anonymize their records, and make them publicly available for statistical analyses. The problem is that Alice's data may be recorded in several statistical databases, one owned by Bob, another one by Carol, another one by Elizabeth, and Frank; and Alice's data may be anonymized differently in each case: Bob may omit Alice's address, Carol her date of birth, Elizabeth her phone number. Now Dave may link Alice's records in Bob's and Carol's databases by the phone number, in Carol's and Elizabeth's databases by the address, etc. He may then rearrange data according to a common ordering, and align the anonymized records so that the gap on each of them is filled by the date in another one. This is a special case of the fusion operation operation described in Sec.~\ref{sec:fusion}. On the other hand, the attack is clearly an instance of the privacy protocol  in Fig.~\ref{fig:crossref}. Re-identifying someone from statistical databases is similar to borrowing the pancake ingredients from a neighbor, and to checking an alibi.

Large-scale versions of the cross-sharing protocol are nowadays routinely launched  as level-above attacks over many services and applications. E.g. the SNet protocol from Fig.~\ref{fig:social} is often sublimated into an instance of a cross-sharing attack from Fig.~\ref{fig:crossref}: providers leverage requests for ongoing access to private channels against users' requests for a single service. In a much publicized incident, a psychology lecturer from Cambridge University, let us call him Dave, designed a Facebook app to collect user data for research purposes \cite{Kosinski:traits,KoganA}, which offered a simple personality quiz, in exchange for full access to users' contact lists. The quiz was taken by 270,000 users, some of them called Bob or Carol, who unwittingly delivered the profiles of 87,000,000 of their contacts, some of them called Alice. The harvested profiles were sold to a political consultancy, Cambridge Analytica \cite{Cambridge-Analytica}, and were used to a great effect in  a level-above instance of the SInf protocol from Fig.~\ref{fig:sublime}.

\section{From security protocol analysis\\ to privacy protocol analysis
%Outlook: Sublimation and deceit
%Monetizing and weaponizing privacy
%Policies and protocols for privacy transformation
}
\label{Sec:Meas}
% !TEX root = 0-PrivT.tex

Security is a useful property. Security requirements are usually publicly declared, with a clear utility. Security protocols implement explicit security requirements. Security protocol failures are unintended protocol runs missed by the protocol designers, usually because of the complexities of the underlying distributed algorithms. The goal of security protocol analysis is to outrun the attackers in detecting and eliminating the unintended runs.

Privacy is a right. But my right to privacy may counter your right to privacy. Private rights need to be balanced against one another, and the privacy requirements are not always agreed upon, or publicly declared. Privacy protocols often implement implicit privacy requirements, sublimated to their level-above deployments. In the realm of privacy, the task of protocol analysis is not any more just to detect the unintended failures with respect to declared requirements.
% or the unintended protocol runs that were declared as undesired
In the realm of protocols that implement private utility, the task of protocol analysis is also to detect the intended sublimated protocol runs that arise from undeclared protocol requirements. 
%Protocol analysis should also analyze the social utility of the undeclared privacy properties of the actually implemented protocols.
Privacy protocol analysis thus provides a technical underpinning for the process of balancing information and value distributions in network society \cite{BenklerY:book,vanDijk:network,zuboff2019age}. 

Given a security protocol, we strive to prove that its declared security requirements are enforced, or to uncover any attacks that may exist. Given a privacy protocol, the task is to establish whose privacy requirements it implements. Alice's and Bob's privacy requirements often clash, and the boundary between protocols and attacks is blurred. At the level above, though, instead of attacks, there are now deceptions to be uncovered and analyzed.

\bibliography{PrivT-ref,PavlovicD,math,philosophy,privacy,security,CT,networks,crypto}
\bibliographystyle{plain}

\appendix

\section*{Appendix: Proofs}
% !TEX root = 0-PrivT.tex

\bprf{ of Prop.~\ref{thm:majorization}}
 Given a doubly substochastic matrix $M$, form the doubly stochastic matrix $\widetilde{M}$ as follows. For any $x = \left( x_1 \cdots x_n\right)$, let $\text{diag}(x)$ be the square matrix with $x$ along its main diagonal, and zeros elsewhere.  
%% Only fixed the first two sentences as an example for Jason. PROOFREAD THE REST. -- d 
Let $\bar{1}$ denote the all $1$'s vector. Then we may define a doubly stochastic matrix.  \[\widetilde{M}= \begin{bmatrix} 
M & I - \text{diag}(M \bar{1}) \\
I- \text{diag}( M^\intercal \bar{1})& M^\intercal
\end{bmatrix} \]
$\ref{stat1} \Rightarrow \ref{stat2}$ Suppose $\ssou = D\gamma$, where $\ssou, \gamma \in \Dis \Sou$, and $D$ is a $n\times n$ doubly substochastic matrix with  $n = \max(\size\ssou,\size\gamma)$ . Then as above we form $\widetilde{D}$ which is a $2n \times 2n$ doubly stochastic matrix. We apply Birkhoff's Theorem \cite{birkmaj}, which states that the set of doubly stochastic matrices is the convex hull of the permutation matrices, to $\widetilde{D}$. Thus we write $\widetilde{D}= \sum_{i=0}^m \lambda_i P'_i$, where $m \lt (2n)!$, $\lambda_i\in [0,1]$, $\sum_{i=0}^{m} \lambda_i \leq1$, and $P'_i$ are $2n \times 2n$ permutation matrices. We may extract the upper left $n\times n $ matrices from each $P'_i$ and denote them by $P_i$. Then we have $\ssou = D\gamma= \sum_{i=0}^m \lambda_i P_i \gamma$, and we are done.
 
$\ref{stat2} \Rightarrow \ref{stat1}$ In a similar manner as above, suppose $\ssou =  \sum_{i=0}^m \lambda_i P_i \gamma$, where the $P_i$ are partial permutations and $\lambda_i$ and $m$ satisfy the conditions above. Then for each $P_i$ we form $\widetilde{P}_i$ which are permutation matrices since they are doubly stochastic with only $1$'s and $0$'s as entries. By Birkhoff's Theorem, $\sum_{i=0}^m \lambda_i \widetilde{P}_i$ is a doubly stochastic matrix, say $D'$, since it is a convex combination of permutation matrices. Extract the upper left $n\times n$ submatrix of $D'$ and label it $D$. Then we have \[\ssou =  \sum_{i=0}^m \lambda_i P_i \gamma = D \gamma. \]
Thus we have $\ref{stat2} \Leftrightarrow \ref{stat1}$.

$\ref{stat1} \Rightarrow \ref{stat3}$ As above, suppose $\ssou = D\gamma$, and let $d_{i,j}$ denote the entry in row $i$, column $j$. We follow the proof of theorem A.4 in \cite{marshallmajor}. Let $P, Q$ be any permutation matrices such that $P \ssou = \ssou^\downarrow$ and $ Q\gamma= \gamma^\downarrow$, i.e., they are reordered in decreasing order. Since $P$ and $Q$ are invertible, $\ssou = D\gamma$ if and only if $P\ssou =P D Q^{-1}Q \gamma$. In addition, $PDQ^{-1}$ is a doubly substochastic matrix since we may simply multiply on the right (left) by $\bar{1}$ $(\bar{1}^\intercal) $ respectively and observe the entries of the resulting vector are in $[0,1]$. Thus we may assume for simplicity that $\ssou$ and $\gamma$ are in decreasing order.
Then we have
\[\jnt\ssou(k)  =  \sum_{i=0}^{k} \ssou({i}) =  \sum_{i=0}^{k} \sum_{j=0}^{n-1}\gamma({j}) d_{i,j}= \sum_{j=0}^{n-1}\gamma({j})\sum_{i=0}^{k}  d_{i,j}. \]
Let $t_j=\sum_{i=0}^{k}  d_{i,j}$ and $\epsilon = k+1- \sum_{j=0}^{n-1} t_j$. Since $D$ is substochastic, it follows that $0 \leq \epsilon \leq k$.
Now we show $\ssou \prec \gamma$.
\begin{align*}
\sum_{i=0}^{k} \ssou({i}) - \sum_{i=0}^{k} \gamma({i})=& \sum_{j=0}^{n-1}\gamma({j})\sum_{i=0}^{k}  d_{i,j} -\sum_{i=0}^{k} \gamma({i}) \\  
=& \sum_{j=0}^{n-1}\gamma({j})t_j -\sum_{i=0}^{k} \gamma({j}) +\gamma(k)\left( k+1- \epsilon - \sum_{j=0}^{n-1} t_j \right) \\
=& \sum_{i=0}^{k} \left(\gamma({i}) - \gamma({k})\right)\left(t_i -1\right) +\sum_{i=k+1}^{n-1} \left(\gamma({i})  -\gamma(k) \right)t_i -\gamma(k)\epsilon \\
\leq& 0
\end{align*}
The last inequality follows since $\gamma({i}) - \gamma({k})$ is $\geq 0$ for $i \leq k$ and $\leq 0$ otherwise. In addition, $0\leq t_i \leq 1$. Thus  $\ssou \prec \gamma$.

$\ref{stat3} \Rightarrow \ref{stat1}$
As before, we may assume $\beta$ and $\gamma$ are in descending order. We now will find where they differ and transform $\gamma$ to $\ssou$ by a series of substochastic transformations and permutations. 

We now perform the first step. Let $j$ be the greatest index where $\ssou(j) \lt \gamma(j)$, and if $j$ does not exist then we are done since they are equal. Let $k\gt j$ be the smallest index such that  $\ssou(k) \gt \gamma(k)$. If $k$ does not exist, i.e., $\ssou$ is less than $\gamma$ component-wise, then form \[ D=\text{diag}( \frac{\ssou(0)}{\gamma(0)},    \frac{\ssou(1)}{\gamma(1)},\dots , \frac{\ssou(m-1)}{\gamma(m-1)} ,0,\dots, 0) \]  
where $\gamma(m)= 0$ and $\gamma(i) \gt 0$ for $i\lt m$.
Clearly $D$ is substochastic and $\ssou= D \gamma$.  
Let us assume $k$ exists. We now define a T-transformation as follows. Let $\lambda \in [0,1]$ and $Q$ a permutation matrix. Then a $T$-transformation has the form \[T= \lambda I + (1-\lambda ) Q.\] 
Observe that \[ \gamma(j) \gt \ssou(j) \geq \ssou(k) \gt \gamma(k). \] 
Now define \[\lambda= 1 - \frac{\min( \gamma(j)-\ssou(j), \ssou(k)- \gamma(k) )}{\gamma(j) -\gamma(k) }.\]
Let $Q$ be the permutation matrix switching out the $j$ and $k$ coordinates, and let $T$ as above. Then let $\gamma'= T\gamma$. Observe that $\ssou \prec \gamma'$ since the total sum is preserved up to 
$j$ since $\jnt\gamma'(i)=\jnt\gamma(i) \geq \jnt\ssou(i)$ for $i\lt j$. At $j$, if 
\[\gamma(j)-\ssou(j) \leq \ssou(k)- \gamma(k), \] 
then 
\[\gamma'(j)= \ssou(j),\] 
otherwise 
\[\gamma'(j)= \gamma(k) + \gamma(j) -\ssou(k).\]
In either case we have $\gamma'(j) \geq \ssou(j)$. Thus $\jnt\gamma'(j)\geq \jnt\ssou(j)$. For $j\lt i \lt  k$, by choice of $i,j$ we have $\gamma'(i)=\gamma(i)= \ssou(i)$, and hence $\jnt\gamma'(i)\geq \jnt\ssou(i)$. Finally for $i\geq k$, by definition of the transformation, $T$, $\jnt\gamma'(i)=\jnt\gamma(i)= \jnt\ssou(i)$. Thus we have $\ssou \prec \gamma'$.

Note that each time we perform this step, since either the $\gamma'(j)=\ssou(j)$ or $\gamma'(k) =\ssou(k)$, we convert one point a $\gamma'$ to a point of $\ssou$. Thus within $n$ steps we will finish, and we are done.

\epr

\bprf{ of Prop.~\ref{prop:join-meet}} We show that there is a common ordering for $\Dmeet B$. It follows that  $\Dmeet B$ is then majorized by the elements of $B$. % The proof for $\Djoin B$ is analogous. --- IT IS FAR FROM OBVIOUS THAT THE PROOF IS ANALOGOUS, SO I AM COMMENTING OUT THIS SENTENCE.

Let us first show  $\Dmeet B$ is ordered by $\ssig$.  The essential condition we must show is that $\Dmeet B \ssig(j) \geq \Dmeet B \ssig(j+1)$. For ease of notation let $\ssou_j$ denote the element of $B$ that minimizes the sum $\sum_{i=0}^j \ssou\ssig(i)$.
\begin{align*}
\Dmeet B \ssig(j) &=  \Big(\partial \bigwedge_{\ssou \in B} \jnt\ssou\ssig \Big)(j) \\
&= \min_{\ssou \in B} \sum_{i=0}^j \ssou\ssig(i) -\min_{\ssou \in B} \sum_{i=0}^{j-1} \ssou\ssig(i) \\
&\geq  \sum_{i=0}^j \ssou_j\ssig(i) -  \sum_{i=0}^{j-1} \ssou_j\ssig(i) \\
&= \ssou_j\ssig(j) \\
&\geq  \ssou_j\ssig(j+1) \\
&= \sum_{i=0}^{j+1} \ssou_j\ssig(i) -  \sum_{i=0}^{j} \ssou_j\ssig(i) \\
&\geq \min_{\ssou \in B} \sum_{i=0}^{j+1} \ssou\ssig(i) -\min_{\ssou \in B} \sum_{i=0}^{j} \ssou\ssig(i) \\
&= \Dmeet B \ssig(j+1)
\end{align*}
Thus $\Dmeet B$ is ordered by $\ssig$, and we now show that it is majorized by all $\gamma \in B$. 
\begin{align*}
\jnt\Dmeet B^\downarrow{(k)} & = \jnt\Dmeet B \ssig {(k)}\\
=& \jnt \Big(\partial \bigwedge_{\ssou \in B} \jnt\ssou\ssig \Big)(k) \\
=& \min_{\ssou \in B} \sum_{i=0}^{k} \ssou\ssig(i) \\
\leq &  \sum_{i=0}^{k} \gamma\ssig(i) \\
=& \jnt\gamma^\downarrow{(k)}
\end{align*}
Observe by definition the bound is tight, since for every $k$ there is a $\ssou \in B$ such that $\jnt\Dmeet B^\downarrow{(k)}=  \jnt\ssou^\downarrow{(k)}$.
\epr

\bprf{ of Lemma ~\ref{lem:cclass}}
By definition, we have $\bigcup\limits_{y\in \Sou} B_y = \Sou$. Now we show, given $u,v \in \Sou$ that $B_u \cap B_v$ is either empty or $B_u=B_v$. If $u \neq v$ and $|B_v|=1$ or $ |B_u|=1$  then $B_u \cap B_v=\emptyset$ since a consistency class with only one element implies that the element can not be in any $\lhd$-cycles.  Let us assume both $B_u$ and $B_v$ have more than one element and $\exists \; w\in B_u \cap B_v$. The previous statements imply $\exists \; \delta_0, \delta_1 , \dots ,\delta_m, \gamma_0,\gamma_1, \dots, \gamma_n \in B$ and $u_1,u_2,\dots,u_m, v_1,v_2,\dots,v_n \in \Sou$ such that we have the two following $\lhd$-cycles
\begin{align*}
	w & \prref{\delta_0} u_1 \dots u_m \prref{\delta_m} w\\
	w & \prref{\gamma_0} v_1 \dots v_n \prref{\delta_n} w
\end{align*}
where there are $i,j$ so that $u_i=u$ and $v_j=v$. Without loss of generality, let $x \in B_u$. We will show $x \in B_v$. Then as above, $x$ is in a $\lhd$-cycle \[x  \prref{\alpha_0} u'_1 \dots u'_p \prref{\alpha_p} x\]
where we there is a $k$ so that $u'_k=u$. Then we may construct the following $\lhd$-cycle\[ x  \prref{\alpha_0} u'_1 \dots  \prref{\alpha_k}u'_k=u_i \dots u_m \prref{\delta_m} w \prref{\gamma_0} v_1 \dots v_n \prref{\delta_n} w \prref{\delta_0} u_1 \dots u_i=u'_k \dots u'_p \prref{\alpha_p} x\]
Thus we have shown that $x \in B_v$.
For the second part of the lemma, we must show that if $B_u \cap B_v = \emptyset$ then $B_u \pref{\delta}B_v$ or  $B_v \pref{\delta} B_u$ for all $\delta \in B$. If there exists $\delta, \gamma \in B$ and $x \in B_u$, $y\in B_v$ so that $x \prref{\delta} y$ and $y \prref{\gamma} x$, then it follows $x \in B_v$ and $y \in B_u$ resulting in a contradiction.
\epr
\bprf{ of Prop.~\ref{prop:imposing-consistency}}
We first show $\widehat\ssou$ is consistent with all elements of $B$. Let $\gamma \in B$. Suppose $u \prref{\gamma} v$. Then either $B_u \cap B_v= \emptyset$ or $B_u = B_v$. If the intersection is empty, then by lemma ~\ref{lem:cclass} it follows $B_u \pref{\ssou}B_v$. Hence $u \pref{\widehat\ssou} v$. It the intersection is nonempty, it follows $B_u=B_v$ and trivially $u \indif{\widehat\ssou} v$.

Conversely, suppose $u \prref{\widehat\ssou} v$. Then $B_u \neq B_v$ and by lemma  ~\ref{lem:cclass} the intersection is empty. In addition, since $w \prref{\ssou} x$ for some $w\in B_u$ and $x \in B_v$ it follows $u \pref{\gamma} v$.
Thus $\widehat\ssou$ is consistent with the elements of $B$.

Now we show $\widehat\ssou$ is maximal with regards to all sources consistent with the elements of $B$ and majorized by $\ssou$. Note trivially that $\widehat\ssou \prec \ssou$ since by definition $\widehat\ssou(u) \leq \ssou(u)$ for all $u\in \Sou$. Suppose $\alpha \in \Dis\Sou$ is consistent with the elements of $B$.
We first make a quick observation, which is that any source consistent with each element of $B$, will be constant on the consistency classes of $B$, i.e., $B_u$ for $u \in \Sou$. Let $w,x \in B_u$ and suppose $w \prref{\alpha} x$. Then since $\alpha$ is consistent with $B$, we have $w \pref{\delta} x$ for all $\delta \in B$. This is a contradiction, since there must exist a $\gamma$ such that $x \prref{\gamma} w$, otherwise $x,w \notin B_u$. From here it is immediate that $\widehat\ssou$ was chosen maximally. This follows since both $\alpha$ and $\widehat{\ssou}$ are constant on consistency classes of $B$. Furthermore since $\alpha \prec \ssou$, it follows for any $u$ that \[\alpha(u) \leq \bigwedge_{  y \in B_u }  \ssou(y) = \widehat \ssou(u)  \] .
Thus $\widehat{\ssou}$ is maximal with regards to the majorization preorder.

The final part of the proposition is to show that the set  $\widehat B = \left\{\widehat \ssou\ |\ \ssou \in B\right\}$ is consistent. It follows from lemma~\ref{lem:cclass} and the previous remark, that any source consistent with each element of $B$ will be constant on the consistency classes. In particular, the lemma implies that the consistency classes are ordered under $\pref{B}$. The previous remark implies that we may find a common ordering based upon the ordering of the consistency classes. Then since we have a common ordering for $\widehat{B}$, by prop.~\ref{prop:cons-order} it is consistent.
\eprf

\end{document}